\pdfoutput=1
\documentclass[jou,american]{apa7}

\usepackage{amsmath}
\usepackage{amssymb}
\usepackage{algorithm}
\usepackage{algpseudocode}
\hypersetup{hidelinks}
\usepackage{csquotes}
\usepackage{float}
\usepackage[style=apa,sortcites=true,sorting=nyt,backend=biber]{biblatex}
\DeclareLanguageMapping{american}{american-apa}
\usepackage{graphicx}
\usepackage{setspace} % Needed for \singlespacing

\usepackage{booktabs} % Needed for \toprule
\usepackage{threeparttable}
\usepackage{diagbox}
\usepackage{adjustbox}

\usepackage{lscape}
\usepackage{multirow}
\begin{document}

\title{Recovering Latent Structures after Variational Bayesian Variable Selection: Fit Assessment and Factor-Number Selection in Partially Exploratory Factor Analysis}

\shorttitle{Post-Selection Fit in PEFA}
\leftheader{Chen and Jin}

\author{Jinsong Chen\textsuperscript{a}		\footnote{Corresponding author: jinsong.chen@live.com} and Yi Jin\textsuperscript{b}}
		
\authorsaffiliations{\textsuperscript{a}Faculty of Education, The University of Hong Kong\\
\textsuperscript{b}Department of Curriculum and Instruction, The Education University of Hong Kong}

\abstract{In partially exploratory factor analysis (PEFA), the loading structure and the number of factors are only weakly specified, and the regularized variational approximation for partially confirmatory factor analysis (PCFA-VA) recovers the structure by Bayesian variable selection: spike-and-slab priors with variational inference attach an inclusion probability to every unspecified loading. This research develops the post-selection assessment that such variable selection requires. We convert the converged solution into a covariance model under hard selection, which thresholds the inclusion probabilities into a fixed sparse pattern, or soft selection, which retains them as weights and yields an effective number of parameters in the spirit of Bayesian and generalized degrees of freedom. We derive the corresponding parameter counts, degrees of freedom, absolute fit diagnostics (RMSEA, SRMR, CFI, TLI), and relative criteria (AIC, BIC, and the algorithm-native evidence lower bound, ELBO). To choose the number of factors we propose a scale-free gain rule with a sustained-drop guard, which retains a factor only when its marginal gain in a criterion exceeds a fixed fraction of the largest gain, and we give conditions under which it exactly recovers a bracketed true dimensionality. Two simulation studies show that the absolute indices track loading recovery and flag under-factoring, that the raw criteria over-factor whereas the gain rule recovers the true dimensionality across a wide range of its threshold, and that the ELBO gain is the most robust variant under heavy nuisance. An empirical example with the 100-item PID-5 shows that the selected solution fits better than the confirmatory 25-facet model even under a plug-in evaluation that disfavors it, that two fully disjoint specifications recover the major latent structure concordantly, and that the specification controls the resolution of mutually collinear factors.}

\maketitle

\section{Introduction}

Partially confirmatory factor analysis (PCFA) was proposed to bridge the practical gap between exploratory factor analysis (EFA), where little prior knowledge is imposed, and confirmatory factor analysis (CFA), where the loading matrix is strongly specified \parencite{chen2021partially}. The regularized variational approximation for PCFA (PCFA-VA) further improves this framework by replacing computationally intensive MCMC estimation with a deterministic variational optimization scheme and by using stochastic search variable selection (SSVS) to identify active and inactive unspecified loadings \parencite{jin2025regularized}. In PCFA-VA, entries of the specification matrix $\mathbf{Q}$ may be required, unrequired, or unspecified, allowing different amounts of substantive knowledge to be incorporated in the same model. Recovering a partially specified loading structure is thus, at its core, a Bayesian variable-selection problem in a high-dimensional latent variable model: each unspecified loading is a candidate variable, the spike-and-slab prior supplies the selection mechanism, and the converged variational posterior attaches an inclusion probability to every candidate.

In this research, we expand the PCFA-VA machinery to accommodate partially exploratory factor analysis (PEFA), where the loading structure and, potentially, the number of factors are partially, and even weakly, specified. The algorithm, the notation, the priors, the variational posterior moments, and the loading activeness rules all remain consistent with the original PCFA-VA formulation. However, the problem encountered in PEFA is substantially different, because $\mathbf{Q}$ contains many unspecified entries and candidate values of $K$ may be compared, which calls for an explicit fitting strategy and accompanying fit statistics.

The purpose of this paper is to equip this variable-selection machinery with the post-selection assessment tools that its use in practice requires: fit statistics analogous to those used in regular SEM and factor analysis, defined so that they respect the selection step that produced the model. These statistics are needed for two related tasks. First, after fitting a model with a fixed $K$ and a partially specified $\mathbf{Q}$ matrix, researchers need absolute fit diagnostics to evaluate whether the selected covariance structure adequately reproduces the observed covariance matrix. Second, when $K$ or $\mathbf{Q}$ varies across candidate models, researchers need relative fit criteria that balance covariance reproduction against model complexity, where the complexity of a selected model is itself nontrivial: a thresholded pattern yields a discrete parameter count, whereas the inclusion probabilities themselves yield an \emph{effective} number of parameters in the spirit of Bayesian and generalized degrees of freedom \parencite{spiegelhalter2002bayesian,efron2004estimation}. The sections below define hard-selected and soft-selected covariance models, their corresponding parameter counts and degrees of freedom, and fit indices based on these quantities, together with a scale-free gain rule for selecting the number of factors and a formal condition under which the rule recovers a bracketed true dimensionality. Two simulation studies calibrate these tools, and an empirical analysis of the Personality Inventory for DSM-5 both demonstrates them and probes their robustness to the specification.

\section{Analytical Framework}

\subsection{The Logic of PEFA with Fit Statistics}

In psychometric analysis, dimensional uncertainty is often unavoidable when there is substantial nuisance structure, such as minor factors or local dependence. This issue is especially salient when the number of latent dimensions is large. The issue is often approached statistically as different retention methods can suggest different numbers of factors. Examples include parallel analysis based on the mean random eigenvalue or the 95th percentile \parencite{horn1965rationale}, the empirical Kaiser criterion \parencite{braeken2017empirical}, and the Hull method \parencite{lorenzoseva2011hull}.

What is often overlooked is the substantive perspective: whether a small factor is ``minor'' or substantively meaningful can depend on the theory, measurement purpose, and intended interpretation \parencite{auerswald2019determine}.

This suggests that there may not always be a single optimal structure of latent dimensions. In semi-open or partially exploratory settings, it may be more realistic to work with a plausible window of factor numbers and focus on whether the major latent pattern is stable and interpretable across that range.

When equipped with both absolute and relative fit statistics, the PCFA-VA framework is well suited to this form of model uncertainty. Absolute fit like RMSEA, SRMR, CFI and TLI can help determine whether a selected major latent pattern is adequate for a given dimensionality, whereas relative fit like AIC and BIC can compare candidate solutions across different values of $K$ or different partially specified $\mathbf{Q}$ matrices. Specifically, we propose to search for the major structure over a window of factors based on a backbone specification matrix $\mathbf{Q}$. The factor window can come from a combination of factor-extraction methods and substantive grounds, and the backbone $\mathbf{Q}$ can be partially specified from substantive knowledge. However, the definitions and behaviors of various fit statistics are not obvious and need to be addressed under the PEFA settings with a window of factors.

\subsection{Fit Assessment after Regularized Variational PCFA}

Let $\mathbf{Y}=(y_{ij})_{N\times J}$ denote the centered observed response matrix for $N$ respondents and $J$ observed variables. Following the PCFA-VA formulation of \textcite{jin2025regularized}, the measurement model for variable $j$ is
\begin{equation}
    \mathbf{Y}_j=\boldsymbol{\eta}\boldsymbol{\Lambda}_j^{\top}+\boldsymbol{\epsilon}_j,
    \qquad
    \boldsymbol{\epsilon}_j\sim N_N(\mathbf{0},\psi_{jj}\mathbf{I}_N),
\end{equation}
where $\mathbf{Y}_j=(y_{1j},\ldots,y_{Nj})^{\top}$, $\boldsymbol{\eta}=(\boldsymbol{\eta}_1,\ldots,\boldsymbol{\eta}_N)^{\top}$ is the $N\times K$ latent factor matrix, and $\boldsymbol{\Lambda}_j=(\lambda_{j1},\ldots,\lambda_{jK})$ is the loading vector for variable $j$. Individual factor scores follow
\begin{equation}
    \boldsymbol{\eta}_i\mid \boldsymbol{\Phi}\sim N_K(\mathbf{0},\boldsymbol{\Phi}),
\end{equation}
where $\boldsymbol{\Phi}=(\phi_{kk^{\prime}})_{K\times K}$ is the factor covariance matrix. As in PCFA-VA, factor scale is identified by standardizing $\boldsymbol{\Phi}$ to a correlation matrix rather than by fixing one loading per factor. The error covariance matrix is diagonal,
\begin{equation}
    \boldsymbol{\Psi}=\operatorname{diag}(\psi_{11},\ldots,\psi_{JJ}).
\end{equation}
Thus, the covariance model implied by a given loading matrix $\boldsymbol{\Lambda}=(\boldsymbol{\Lambda}_1,\ldots,\boldsymbol{\Lambda}_J)^{\top}$ and parameter vector $\boldsymbol{\theta}$, which collects the free loadings, the factor correlations, and the error variances, is
\begin{equation}
    \boldsymbol{\Sigma}(\boldsymbol{\theta})
    =
    \boldsymbol{\Lambda}\boldsymbol{\Phi}\boldsymbol{\Lambda}^{\top}
    +
    \boldsymbol{\Psi}.
\end{equation}
If a different identification scheme is adopted, such as fixing one loading per factor and estimating factor variances, the parameter counts and degrees of freedom below should be modified accordingly.

The partially specified loading structure is encoded by $\mathbf{Q}=(q_{jk})_{J\times K}$. Consistent with PCFA-VA, $q_{jk}=1$ denotes a required loading, $q_{jk}=0$ denotes an unrequired loading fixed at zero, and $q_{jk}=-1$ denotes an unspecified loading. For each unspecified loading, PCFA-VA introduces a latent loading activeness indicator $\gamma_{jk}$ through the SSVS prior
\begin{equation}
    \lambda_{jk}\mid \gamma_{jk}
    \sim
    N\{0,(1-\gamma_{jk})v_0+\gamma_{jk}v_1\},
    \qquad
    \gamma_{jk}\sim \operatorname{Bernoulli}(\rho),
\end{equation}
with $v_1>v_0>0$, so that an active loading is drawn from the slab with the larger variance $v_1$ and an inactive loading from the spike with the small variance $v_0$. Required loadings ($q_{jk}=1$) instead receive a diffuse normal prior $\lambda_{jk}\sim N(0,h_0)$ with a large fixed variance $h_0$ and are therefore freely estimated. Let $\boldsymbol{\Gamma}$ collect the indicators $\gamma_{jk}$ for the unspecified entries. Here $\rho$ denotes the Bernoulli inclusion probability in the SSVS prior, as in Eq.~(6) of \textcite{jin2025regularized}; it is often fixed at $0.5$ when no prior loading activeness information is available. To avoid confusion with the entries $q_{jk}$ of $\mathbf{Q}$, let $q_{\mathrm{var}}(\cdot)$ denote the variational density and define
\begin{equation}
    \widehat{\pi}_{jk}=q_{\mathrm{var}}(\gamma_{jk}=1),
    \qquad
    \widehat{\lambda}_{jk}=E_{q_{\mathrm{var}}}(\lambda_{jk})
\end{equation}
as the converged variational posterior loading activeness probability and posterior mean of the loading. Determining whether an unspecified entry is required or unrequired is therefore equivalent to identifying whether the corresponding loading is active. The default PCFA-VA activeness rule is
\begin{equation}
    \widehat{\pi}_{jk}\geq \tau,
\end{equation}
where threshold $\tau=.50$ is typically used, paralleling the rule $\widehat{\pi}_{jk}\geq .50\Leftrightarrow\widehat{\gamma}_{jk}=1$ of \textcite{jin2025regularized}, although sensitivity analyses may consider more conservative values \parencite{rovckova2014emvs,you2023approximated}.

\subsection{Hard and Soft Selection}

For SEM-like fit assessment, the converged variational solution is converted into a covariance model. The hard-selected model treats each unspecified loading as either active or inactive. Define
\begin{equation}
    a_{jk}^{H}(\tau)=
    \begin{cases}
        1, & q_{jk}=1,\\
        0, & q_{jk}=0,\\
        I(\widehat{\pi}_{jk}\geq \tau), & q_{jk}=-1,
    \end{cases}
\end{equation}
and
\begin{equation}
    \widetilde{\lambda}_{jk}^{H}
    =
    a_{jk}^{H}(\tau)\widehat{\lambda}_{jk}.
\end{equation}
Let $\widetilde{\boldsymbol{\Lambda}}_{H}$ be the resulting loading matrix. Let $\widehat{\boldsymbol{\Phi}}$ denote the standardized variational estimate of the factor correlation matrix and $\widehat{\boldsymbol{\Psi}}=\operatorname{diag}\{\widehat{\psi}_{11},\ldots,\widehat{\psi}_{JJ}\}$ denote the variational estimate of the diagonal error covariance matrix. The hard-selected fitted covariance matrix is
\begin{equation}
    \widehat{\boldsymbol{\Sigma}}_{H}
    =
    \widetilde{\boldsymbol{\Lambda}}_{H}
    \widehat{\boldsymbol{\Phi}}
    \widetilde{\boldsymbol{\Lambda}}_{H}^{\top}
    +
    \widehat{\boldsymbol{\Psi}}.
\end{equation}
This hard-selected covariance model is the closest analogue to a conventional sparse CFA/SEM model, because it produces a fixed loading pattern after the PCFA-VA loading activeness step.

The number of unique observed covariance moments is
\begin{equation}
    m=\frac{J(J+1)}{2}.
\end{equation}
Under the correlation-matrix identification of $\boldsymbol{\Phi}$, the nominal number of factor-correlation parameters is $K(K-1)/2$. The hard-selected loading count is
\begin{equation}
    p_{\Lambda,H}(\tau)
    =
    \sum_{j=1}^{J}\sum_{k=1}^{K} I(q_{jk}=1)
    +
    \sum_{j=1}^{J}\sum_{k=1}^{K} I(q_{jk}=-1)I(\widehat{\pi}_{jk}\geq \tau).
\end{equation}
The nominal hard-selected covariance-parameter count is
\begin{equation}
    t_{H}^{\mathrm{nom}}(\tau)
    =
    p_{\Lambda,H}(\tau)+J+\frac{K(K-1)}{2},
\end{equation}
where $J$ counts error variances. If the selected loading pattern is locally identified in the ordinary CFA sense, the nominal hard-selected degrees of freedom are
\begin{equation}
    df_H^{\mathrm{nom}}=m-t_{H}^{\mathrm{nom}}.
\end{equation}

For fully unspecified or weakly specified $\mathbf{Q}$ matrices, nominal counting may overstate model dimension because some rotational directions can remain redundant \parencite{browne2001overview}. A rank-adjusted definition is therefore preferable. Let
\begin{equation}
    \boldsymbol{\Delta}_{H}
    =
    \left.
    \frac{\partial \operatorname{vech}\{\boldsymbol{\Sigma}(\boldsymbol{\theta})\}}
    {\partial \boldsymbol{\theta}_{H}^{\top}}
    \right|_{\widehat{\boldsymbol{\theta}}_{H}}
\end{equation}
be the Jacobian of the unique covariance elements with respect to the hard-selected parameter vector, including selected active loadings, factor correlations, and error variances. Define the local hard-selected dimension as
\begin{equation}
    t_H^{\mathrm{rank}}=\operatorname{rank}(\boldsymbol{\Delta}_{H}).
\end{equation}
For the final hard-selected parameter count, use
\begin{equation}
    t_H=
    \begin{cases}
        t_H^{\mathrm{nom}}, & \text{if the selected pattern is locally identified},\\
        t_H^{\mathrm{rank}}, & \text{if rotational redundancy remains}.
    \end{cases}
\end{equation}
The corresponding hard-selected degrees of freedom are
\begin{equation}
    df_H=m-t_H,
\end{equation}
where $t_H=t_H^{\mathrm{rank}}<t_{H}^{\mathrm{nom}}$ when redundant rotational directions are detected. In empirical reporting, $t_{H}^{\mathrm{nom}}$ is useful as a transparent parameter count, whereas $t_H$ is the preferred complexity term for chi-square-type fit indices and hard-selection relative fit.

Two remarks delimit this treatment of identification. First, classical sufficient conditions can certify local identification of a selected pattern without the numerical rank check: patterns in which each factor retains at least three pure markers, or two pure markers when the factors are correlated, satisfy the row-deletion conditions of \textcite{anderson1956statistical} (see also \cite{bekker1997generic} for generic global results). In the workflows below, the required backbone markers together with the sparsity that hard selection induces typically place the selected pattern in this regime, and the rank check confirmed it whenever it was run. Second, when $J$ is large the Jacobian computation can be slow and the check may be skipped; because $t_H^{\mathrm{rank}}\leq t_H^{\mathrm{nom}}$, substituting the nominal count then \emph{overstates} complexity, which understates $df_H$ and inflates the complexity penalties, so all fit conclusions drawn under the nominal count err on the conservative side.

A soft-selected or effective version can be used as a sensitivity analysis. Define
\begin{equation}
    a_{jk}^{S}=
    \begin{cases}
        1, & q_{jk}=1,\\
        0, & q_{jk}=0,\\
        \widehat{\pi}_{jk}, & q_{jk}=-1,
    \end{cases}
    \qquad
    \widetilde{\lambda}_{jk}^{S}
    =
    a_{jk}^{S}\widehat{\lambda}_{jk}.
\end{equation}
Then
\begin{equation}
    \widehat{\boldsymbol{\Sigma}}_{S}
    =
    \widetilde{\boldsymbol{\Lambda}}_{S}
    \widehat{\boldsymbol{\Phi}}
    \widetilde{\boldsymbol{\Lambda}}_{S}^{\top}
    +
    \widehat{\boldsymbol{\Psi}}.
\end{equation}
The corresponding effective loading count is
\begin{equation}
    p_{\Lambda,S}
    =
    \sum_{j=1}^{J}\sum_{k=1}^{K} I(q_{jk}=1)
    +
    \sum_{j=1}^{J}\sum_{k=1}^{K} I(q_{jk}=-1)\widehat{\pi}_{jk},
\end{equation}
with effective total parameter count and effective degrees of freedom
\begin{equation}
    t_S
    =
    p_{\Lambda,S}+J+\frac{K(K-1)}{2},
    \qquad
    df_S=m-t_S.
\end{equation}
The soft-selected count has a direct interpretation as an \emph{effective number of parameters}. Because each $\gamma_{jk}$ is Bernoulli under the variational posterior, $p_{\Lambda,S}$ is exactly the variational posterior expectation of the number of active loadings, so that $t_S=E_{q_{\mathrm{var}}}\{t_H^{\mathrm{nom}}(\boldsymbol{\Gamma})\}$: the discrete count of the hard-selected pattern, averaged over the selection uncertainty that the posterior retains. It thereby plays the same role as the effective parameter counts of the Bayesian model-comparison literature, $p_D$ in the DIC \parencite{spiegelhalter2002bayesian} and $p_{\mathrm{WAIC}}$ \parencite{watanabe2010asymptotic,gelman2014understanding}, and as the generalized degrees of freedom used to price data-driven selection in regularized regression \parencite{efron2004estimation,zou2007degrees}. This connection also locates the post-selection caveat precisely. The hard-selected $df_H$ is \emph{conditional} on the selected pattern: the data chose which loadings to retain, and that choice consumes information that a fixed prespecified pattern would not, so chi-square-type quantities computed from $df_H$ inherit a degree of post-selection optimism \parencite{berk2013valid}. The gap $t_S-p_{\Lambda,H}(\tau)-J-K(K-1)/2$, equivalently $\sum_{q_{jk}=-1}\{\widehat{\pi}_{jk}-I(\widehat{\pi}_{jk}\geq\tau)\}$, measures how much selection uncertainty the thresholding discards: when every $\widehat{\pi}_{jk}$ is near zero or one the hard and soft counts nearly coincide and the conditional treatment is innocuous, whereas many probabilities near $\tau$ signal that the hard-selected indices should be read with extra caution and checked against their soft counterparts.

Note that the soft-selected $df_S$ should not replace $df_H$ for RMSEA, CFI, or TLI, because those indices are traditionally tied to a fixed selected model in which each loading is either freely estimated or fixed. However, $df_S$ can be useful for evaluating whether relative-fit conclusions depend heavily on the threshold $\tau$ or on the use of an \textit{effective} parameter count under model uncertainty.

\subsection{Absolute and Relative Fit Indices}

Let $\mathbf{S}$ denote the sample covariance matrix and let $N^{\ast}=N-1$ to match the usual covariance-structure SEM scaling. The normal-theory covariance discrepancy evaluated at the hard-selected fitted covariance matrix is
\begin{equation}
    F_H
    =
    \log|\widehat{\boldsymbol{\Sigma}}_{H}|
    +
    \operatorname{tr}(\mathbf{S}\widehat{\boldsymbol{\Sigma}}_{H}^{-1})
    -
    \log|\mathbf{S}|
    -
    J.
\end{equation}
The associated chi-square-type statistic is
\begin{equation}
    T_H=N^{\ast}F_H.
\end{equation}
Because $T_H$ is evaluated at regularized variational estimates and after loading-pattern selection, it should be interpreted as a pseudo chi-square statistic rather than as an exact ML likelihood-ratio test. Two further qualifications sharpen this reading. First, the analyses below standardize the observed variables, so $\mathbf{S}$ is a correlation matrix; because the factor model with free error variances is scale invariant, the discrepancy value is unaffected by the standardization, although the distribution theory for correlation structures differs from the covariance case \parencite{cudeck1989analysis}, one more reason to treat $T_H$ as descriptive. Second, the variational posterior means are shrunken by the SSVS prior rather than chosen to minimize $F_H$ over the selected pattern, so $T_H$ is bounded below by the minimized discrepancy of that pattern; the plug-in indices are in this sense conservative for the selected model.

\paragraph{Absolute fit statistics}
The RMSEA analogue is
\begin{equation}
    \operatorname{RMSEA}_{H}
    =
    \sqrt{
    \max\left(
    \frac{T_H-df_H}{df_HN^{\ast}},
    0
    \right)
    }.
\end{equation}
RMSEA should not be interpreted when $df_H\leq 0$. For SRMR, let $\mathbf{R}$ and $\widehat{\mathbf{R}}_{H}$ denote the sample and hard-selected model-implied correlation matrices. Then
\begin{equation}
    \operatorname{SRMR}_{H}
    =
    \sqrt{
    \frac{2}{J(J+1)}
    \sum_{j=1}^{J}\sum_{\ell=1}^{j}
    (r_{j\ell}-\widehat{r}_{H,j\ell})^2
    }.
\end{equation}
SRMR is a direct standardized residual index and does not depend on a degrees-of-freedom correction.

The independence baseline model is
\begin{equation}
    \widehat{\boldsymbol{\Sigma}}_0=\operatorname{diag}(\mathbf{S}),
    \qquad
    df_0=\frac{J(J-1)}{2}.
\end{equation}
Its discrepancy statistic is
\begin{equation}
    T_0
    =
    N^{\ast}
    \left\{
    \log|\widehat{\boldsymbol{\Sigma}}_0|
    +
    \operatorname{tr}(\mathbf{S}\widehat{\boldsymbol{\Sigma}}_0^{-1})
    -
    \log|\mathbf{S}|
    -
    J
    \right\}.
\end{equation}
The PCFA-VA analogues of CFI \parencite{bentler1990comparative} and TLI \parencite{tucker1973reliability} are
\begin{equation}
    \operatorname{CFI}_{H}
    =
    1-
    \frac{\max(T_H-df_H,0)}
    {\max(T_H-df_H,T_0-df_0,0)},
\end{equation}
\begin{equation}
    \operatorname{TLI}_{H}
    =
    \frac{T_0/df_0-T_H/df_H}
    {T_0/df_0-1}.
\end{equation}
By convention, $\operatorname{CFI}_{H}$ is set to $1$ when the denominator $\max(T_H-df_H,T_0-df_0,0)$ equals zero (both model and baseline fit better than their degrees of freedom), and $\operatorname{TLI}_{H}$ is reported only when $df_H>0$. Although CFI and TLI are incremental indices relative to the independence baseline, they are grouped with the absolute covariance-fit diagnostics because they evaluate the adequacy of a single selected PCFA-VA solution. Conventional cutoff values should be treated as descriptive heuristics rather than strict decision rules in this one-stage regularized VB setting \parencite{west2012model}.

\paragraph{Relative fit statistics}
Because PCFA-VA is estimated by variational optimization, the evidence lower bound (ELBO) is the most direct algorithm-native relative-fit quantity. Let $\mathbf{Z}=(\boldsymbol{\Lambda},\boldsymbol{\eta},\boldsymbol{\Phi},\boldsymbol{\Psi},\boldsymbol{\Gamma},\rho)$ collect the unknown quantities in the PCFA-VA model, with $\rho$ omitted when the SSVS inclusion probability is fixed. For a candidate model indexed by $(K,\mathbf{Q})$, define
\begin{equation}
    \operatorname{ELBO}(K,\mathbf{Q})
    =
    E_{q_{\mathrm{var}}(\mathbf{Z})}
    \left[
    \log p\{\mathbf{Y},\mathbf{Z}\mid K,\mathbf{Q}\}
    \right]
    -
    E_{q_{\mathrm{var}}(\mathbf{Z})}
    \left[
    \log q_{\mathrm{var}}(\mathbf{Z})
    \right],
\end{equation}
so that
\begin{equation}
    \operatorname{ELBO}(K,\mathbf{Q})\leq \log p(\mathbf{Y}\mid K,\mathbf{Q}),
\end{equation}
where larger ELBO values indicate better relative variational evidence. The bound satisfies $\operatorname{ELBO}(K,\mathbf{Q})=\log p(\mathbf{Y}\mid K,\mathbf{Q})-\operatorname{KL}\{q_{\mathrm{var}}(\mathbf{Z})\,\|\,p(\mathbf{Z}\mid\mathbf{Y},K,\mathbf{Q})\}$, so treating ELBO differences as evidence differences implicitly assumes that the variational gap, the KL term, is comparable across the candidates \parencite{blei2017variational}; this assumption cannot be verified directly, and the simulations below probe its practical consequences by scoring the ELBO against the true dimensionality. ELBO comparisons are most meaningful for candidate values of $K$ or competing $\mathbf{Q}$ matrices fit with the same data preprocessing, likelihood constants, priors, and hyperparameters. Because the ELBO is attached to the fitted variational PCFA-VA model before thresholding, it is not generally useful for comparing different threshold values $\tau$ unless the model is refit after imposing the selected loading pattern.

For SEM-like relative fit, one can also evaluate the plug-in log-likelihood at the hard-selected covariance estimate:
\begin{equation}
    \ell_H
    =
    -\frac{N^{\ast}}{2}
    \left[
    J\log(2\pi)
    +
    \log|\widehat{\boldsymbol{\Sigma}}_{H}|
    +
    \operatorname{tr}(\mathbf{S}\widehat{\boldsymbol{\Sigma}}_{H}^{-1})
    \right].
\end{equation}
The hard-selection information criteria are
\begin{equation}
    \operatorname{AIC}_{H}
    =
    -2\ell_H+2t_H,
    \qquad
    \operatorname{BIC}_{H}
    =
    -2\ell_H+\log(N^{\ast})t_H.
\end{equation}
The hard-selected version is recommended as the primary relative-fit approach because it compares fixed sparse covariance models after the PCFA-VA loading activeness step.

For the soft-selected covariance matrix, define
\begin{equation}
    \ell_S
    =
    -\frac{N^{\ast}}{2}
    \left[
    J\log(2\pi)
    +
    \log|\widehat{\boldsymbol{\Sigma}}_{S}|
    +
    \operatorname{tr}(\mathbf{S}\widehat{\boldsymbol{\Sigma}}_{S}^{-1})
    \right].
\end{equation}
The soft-selection information criteria are
\begin{equation}
    \operatorname{AIC}_{S}
    =
    -2\ell_S+2t_S,
    \qquad
    \operatorname{BIC}_{S}
    =
    -2\ell_S+\log(N^{\ast})t_S.
\end{equation}
The soft criteria are not substitutes for the hard-selection criteria. They are sensitivity checks for cases in which several unspecified loadings have inclusion probabilities near the threshold, or they can be used to evaluate the role of \textit{effective} parameter counts as described above.\footnote{Other BIC-type indices, such as SABIC and EBIC, can be computed as supplementary sensitivity criteria. They are not emphasized here because preliminary applications suggest that they add little beyond AIC/BIC or can be overly conservative for weak but interpretable loadings.}

\subsection{A Scale-Free Gain Rule for Selecting the Number of Factors}
Selecting $K$ from any of the relative criteria above raises a common difficulty. At a global optimum of the variational problem the ELBO is nondecreasing in $K$, because the $(K+1)$-factor variational family nests the $K$-factor family, so each added factor can only raise the optimized bound; in practice the coordinate-ascent optimizer attains local optima, so the estimated ELBO path is nondecreasing up to optimization noise and can show small dips past the truth, a point that shapes the rule below. The information criteria, although penalized, keep improving slightly past the truth whenever weak nuisance dimensions are present. The raw optima of all these criteria therefore tend to over-factor. What separates a genuine factor from a nuisance one is not \emph{whether} it improves the criterion but \emph{by how much}: genuine factors produce large improvements, nuisance factors only small ones. This is precisely the logic of the scree test \parencite{cattell1966scree}, retaining factors up to the ``elbow'' of the curve, later given non-graphical, numerical form \parencite{raiche2013nongraphical} and, more generally, automated as knee or elbow detection on a criterion-versus-complexity curve \parencite{salvador2004determining,satopaa2011kneedle}.

We adopt a simple scale-free version of this idea for the PEFA selection path and, because it thresholds the marginal \emph{gain} rather than locating a literal elbow point on a plotted curve, refer to it as the \emph{gain rule}. Let $c(K)$ denote any of the criteria oriented so that larger is better (the ELBO itself, or the negation of an information criterion, $-\!\operatorname{AIC}$ or $-\!\operatorname{BIC}$), and let $g(K)=c(K)-c(K-1)$ be its marginal gain, with $g_{\max}=\max_K g(K)$; we call $g(K)$ the \emph{ELBO gain}, \emph{AIC gain}, or \emph{BIC gain} accordingly. We retain factors up to the point beyond which no further factor contributes more than $\delta$ percent of the largest gain,
\begin{equation}
\widehat{K}_{\delta}=\min\Bigl\{K:\; g(K+1) < \tfrac{\delta}{100}\,g_{\max}\Bigr\},
\qquad \delta=10 .
\label{eq:elbow}
\end{equation}
Because the estimated criterion paths are subject to sampling and optimization noise, a local optimum at some $K$ can produce a transient sub-threshold gain that rebounds immediately, and the first-crossing form of Equation~\ref{eq:elbow} would mistake such a dip for the elbow. The operational rule therefore carries a \emph{sustained-drop guard} with parameter $s$: a sub-threshold gain counts as the elbow only when the following $s-1$ gains are also below $\tfrac{\delta}{100}\,g_{\max}$. We use $s=2$ throughout, in the simulations as well as the empirical analysis. The guard costs $s-1$ extra units of over-factoring headroom in the window but prevents a single noisy dip from terminating the sweep prematurely.

Because the rule compares a \emph{ratio} of gains, it is scale-free: although the absolute size of a step grows with $N\cdot J$, $\widehat{K}_{\delta}$ is invariant to rescaling $c$, so no calibrated absolute threshold is required. Its logic can also be stated exactly. Whenever the candidate window brackets the truth $K^{\ast}$ with $s$ steps of headroom and the gains \emph{separate} at $K^{\ast}$, every gain up to $K^{\ast}$ clearing the threshold and the $s$ gains after $K^{\ast}$ falling below it, the guarded rule returns exactly $K^{\ast}$ (Proposition~1 in Appendix~\ref{app:gainrule}). The separation condition is a statement about the estimated gains, and how often it holds in data, and over how wide a range of $\delta$, is an empirical question that the studies below answer directly; Appendix~\ref{app:gainrule} reports the full sensitivity of the selections to $\delta$ over $[2,40]$. The bracketing requirement is the rule's one structural demand: the window must contain both the large gains of genuine factors and at least one of the small gains of nuisance factors that anchors the comparison. If the window lies entirely below the truth every gain is large and the rule returns the window's upper bound; if it lies entirely above the truth the dominant jump is absent and the rule returns its lower bound. A selection at either boundary is therefore itself informative: it signals a mis-centered window, which one extends upward or downward, respectively, and refits until the selected $K$ falls in the interior. In practice the window is centered on a cheap initial estimate (e.g., parallel analysis) and widened as needed; the studies below show that a $\pm1$ window already suffices when the estimate is well centered, whereas a $\pm3$ window provides insurance against a poorly centered one.

The gain rule can thus be applied to the ELBO, AIC, or BIC alike, and which one is preferred is an empirical question. Simulation Study One shows that in low-dimensional, low-nuisance settings the three gain variants agree and all recover the true $K$; Simulation Study Two shows that under heavy nuisance they separate, and the ELBO gain, computed on the algorithm-native, monotone bound, is the most accurate.

\section{Simulation Study}

Simulation Study One below calibrates the proposed diagnostics under a fully realized partially exploratory workflow, and Simulation Study Two extends them to a high-dimensional setting that matches the empirical inventories analyzed afterward.

\subsection{Simulation Study One}

\subsubsection{Design}

\paragraph{Population model.}
Data were generated from a factor model with $K=5$ correlated factors and
$J=K\times\mathrm{ipf}$ continuous indicators, where the number of items per
factor was $\mathrm{ipf}\in\{4,8\}$ (so $J=20$ or $40$). Each indicator had a
primary loading of $.6$ on its target factor. To create realistic specification
uncertainty, every factor also carried $\mathrm{cpf}=\mathrm{ipf}/2$ secondary
(cross-) loadings of magnitude $.2$ with alternating signs on a neighboring
factor. Inter-factor correlations were homogeneous at $\phi\in\{.25,.5\}$. Two
nuisance conditions were crossed in: \emph{none}, with no additional structure,
and \emph{moderate}, which added a diffuse minor factor that loaded weakly on all
items (loadings drawn from $U(.09,.11)$ with alternating signs and uncorrelated
with the major factors), emulating the minor structure that motivates a partially
exploratory analysis. Sample size was $N\in\{500,1000\}$. Crossing $N$,
$\mathrm{ipf}$, $\phi$, and the nuisance factor produced $2\times2\times2\times2
=16$ conditions, each replicated $200$ times. All variables were standardized, so
the sample and model-implied matrices are correlation matrices.

\paragraph{Estimation and specification.}
Because the number of factors is treated as uncertain, each data set was first
passed through six factor-retention rules: parallel analysis using the mean
($\mathrm{PA}_m$) and the $95$th-percentile ($\mathrm{PA}_{.95}$) reference
eigenvalue, the Kaiser criterion (KC), the empirical Kaiser criterion (EKC), the
Hull method, and a sequential model test (SMT), to obtain a candidate
dimensionality $\widehat{K}$. Reflecting the partially exploratory stance,
PCFA-VA was then fit at $\widehat{K}-1$, $\widehat{K}$, and $\widehat{K}+1$,
spanning a plausible range around each rule's estimate. For a fitted model with
$K$ factors, the specification matrix $\mathbf{Q}$ contained a backbone Q-matrix $\mathbf{Q_0}$ with $K_0\in\{.4K,.6K,.8K\}$ partially specified
factors; within each specified factor the first $\mathrm{ipf}/2$ primary loadings
were marked required ($q_{jk}=1$) and all remaining entries were left unspecified
($q_{jk}=-1$). The design thus varies how much structure the analyst supplies,
from $20\%$ to $40\%$ of the major loadings (equivalently $4\%$ to $8\%$ of all
loadings) as $K_0$ ranges from $.4K$ to $.8K$. At convergence, the
variational inclusion probabilities $\widehat{\pi}_{jk}$ were thresholded at
$\tau=.50$ for hard selection and retained as weights for soft selection.

\paragraph{Outcomes.}
Recovery of the major loading pattern was summarized after matching estimated to
true factors up to permutation and reflection, by the root mean square error (RMSE), bias, and the false-negative (FN) and false-positive (FP) rates of major loadings. Absolute fit was evaluated with $\mathrm{RMSEA}_H$, $\mathrm{SRMR}_H$,
$\mathrm{CFI}_H$, and $\mathrm{TLI}_H$, reported both as mean values and as the
percentage of replications within the conventional cutoffs ($\mathrm{RMSEA}\le.06$,
$\mathrm{SRMR}\le.10$, $\mathrm{CFI}\ge.90$, $\mathrm{TLI}\ge.90$). Relative fit was
scored by how often each of $\mathrm{AIC}_H$, $\mathrm{BIC}_H$, $\mathrm{AIC}_S$,
$\mathrm{BIC}_S$, and the ELBO, together with the AIC, BIC, and ELBO gain rules
(Equation~\ref{eq:elbow}, with the sustained-drop guard, $s=2$), selected the
data-generating $K=5$ within the candidate triplet $\{4,5,6\}$. For each
replication and specification level, the triplet was assembled from the fitted
models with $K_{\mathrm{fit}}\in\{4,5,6\}$ across the retention-rule paths; a
complete triplet was available in every replication, for $9{,}600$ scored
triplets in total.

\subsubsection{Factor-number extraction}
\begin{figure*}[tp]
\centering
\includegraphics[width=\linewidth]{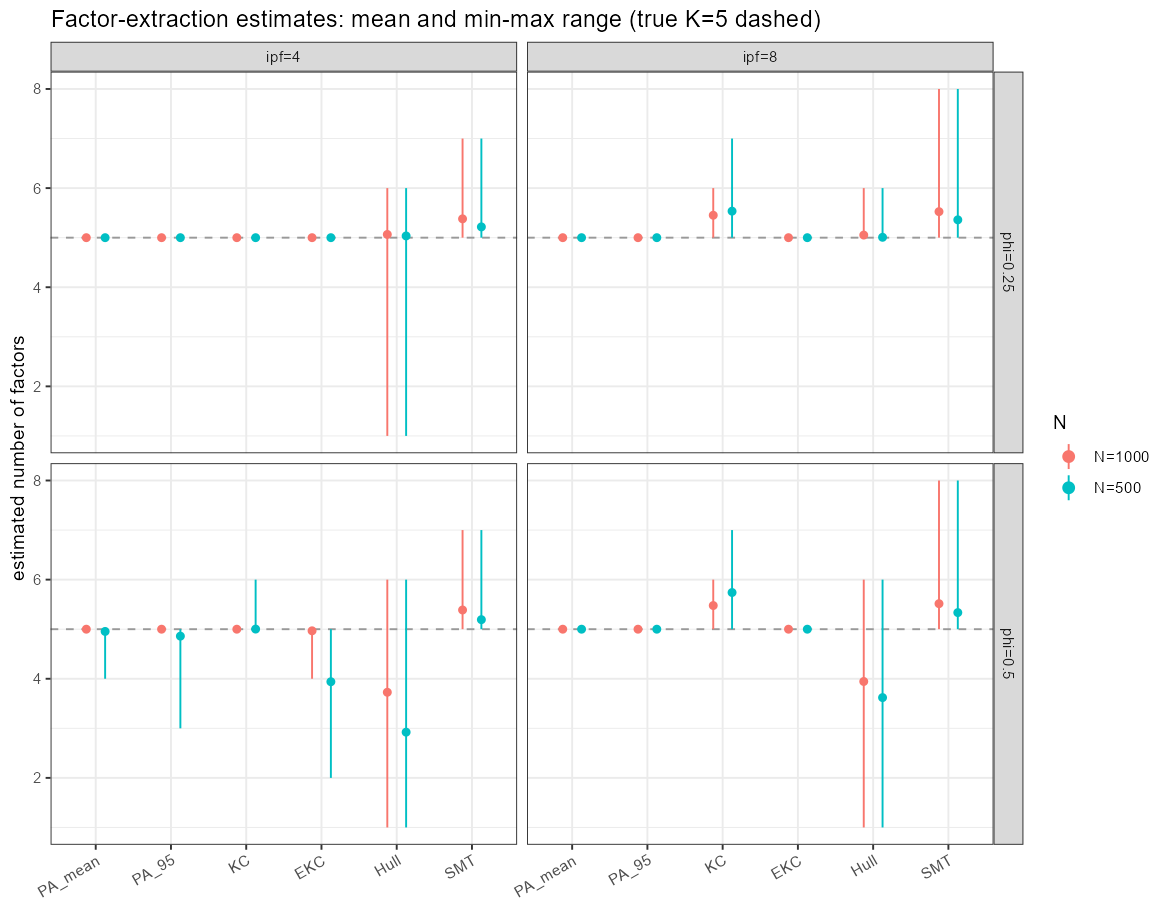}
\caption{Factor-extraction estimates (mean and min--max range across
replications) for each retention rule; the dashed line marks the true $K=5$.
Extraction is the first step of the workflow, seeding the candidate range of $K$.}
\label{fig:extract}
\end{figure*}

Because the workflow begins by estimating the dimensionality, we first examine the
six retention rules that seed the candidate range of $K$
(Figure~\ref{fig:extract}). The parallel-analysis rules were the most accurate
($\mathrm{PA}_m$ recovered $K=5$ in $99.4\%$ of data sets and $\mathrm{PA}_{.95}$
in $98.3\%$), with narrow ranges. The remaining rules were less stable: EKC tended
to under-extract under high factor correlation ($90.2\%$ correct overall), KC and
SMT systematically over-extracted and never selected fewer than five factors
($73.2\%$ and $65.2\%$ correct), and the Hull method was the most erratic, ranging
from one to six factors ($76.1\%$). This disagreement is itself the motivation for
the partially exploratory workflow: rather than trusting any single rule, the
analyst fits a small range of $K$ around the rules' suggestions and uses the fit
machinery to adjudicate.

\subsubsection{Absolute fit and structure recovery}
\begin{figure*}[tp]
\centering
\includegraphics[width=0.9\linewidth]{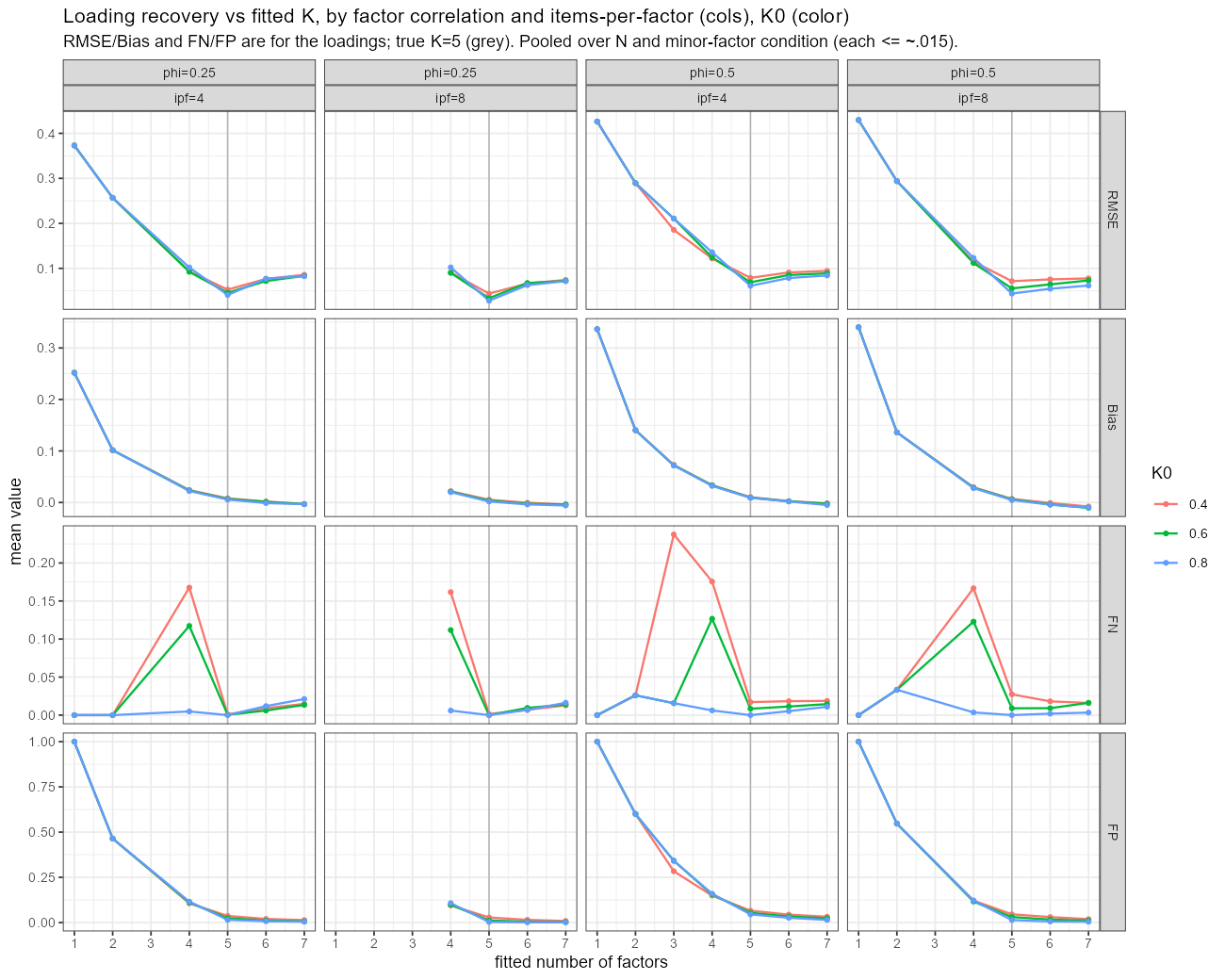}
\caption{Loading recovery versus the fitted number of factors (true $K=5$, grey line), by factor
correlation and items-per-factor (columns) and specification level $K_0$ (color). Pooled
over sample size and the minor-factor condition, each of which shifted recovery by at
most about $.015$.}
\label{fig:fit-trend}
\end{figure*}

Table~\ref{tab:abs-fit} reports absolute fit and loading recovery as a function of the
fitted number of factors, pooled over the 16 conditions and 200 replications, and
Figure~\ref{fig:fit-trend} breaks the recovery out by factor correlation,
items-per-factor, and specification level. Because the candidate range is seeded
by the retention rules rather than by the (unknown) true $K$, the fitted dimensionality
varies more widely than $\{4,5,6\}$.

When the fitted dimensionality matched the truth ($K_{\mathrm{fit}}=5$), loading
recovery was excellent (RMSE $=.052$, near-zero bias) and every absolute index lay
comfortably within its cutoff (mean $\mathrm{RMSEA}_H=.018$, $\mathrm{CFI}_H=.982$,
$\mathrm{TLI}_H=.979$; essentially $100\%$ of replications within all cutoffs).
Thus good loading recovery and good SEM-like fit coincided at the true dimensionality.
The converse held in the under-factored direction: as $K_{\mathrm{fit}}$ fell
below five, recovery and fit deteriorated in lockstep (at $K_{\mathrm{fit}}=4$,
RMSE $=.109$ and $\mathrm{CFI}_H=.881$; at $K_{\mathrm{fit}}=1$, RMSE $=.427$ and
$\mathrm{CFI}_H=.705$). Pooled across fitted models, RMSE correlated
$.77$ with $\mathrm{RMSEA}_H$ and $-.82$ with $\mathrm{CFI}_H$, confirming a
strong concordance between recovery and absolute fit.

\begin{table*}[tp]
\centering
\caption{Absolute fit and parameter recovery by fitted number of factors.}
\label{tab:abs-fit}
\resizebox{\linewidth}{!}{%
\begin{tabular}[t]{crrrrrrrrrrrr}
\toprule
\multicolumn{1}{c}{ } & \multicolumn{4}{c}{Parameter recovery} & \multicolumn{4}{c}{Absolute fit (mean)} & \multicolumn{4}{c}{\% within cutoff} \\
\cmidrule(l{3pt}r{3pt}){2-5} \cmidrule(l{3pt}r{3pt}){6-9} \cmidrule(l{3pt}r{3pt}){10-13}
$K_{\mathrm{fit}}$ & RMSE & Bias & FN & FP & RMSEA & SRMR & CFI & TLI & RMSEA & SRMR & CFI & TLI\\
\midrule
1 & 0.427 & 0.338 & 0.000 & 1.000 & 0.079 & 0.073 & 0.705 & 0.679 & 0.0 & 100.0 & 0.0 & 0.0\\
2 & 0.291 & 0.139 & 0.029 & 0.579 & 0.069 & 0.064 & 0.784 & 0.756 & 0.9 & 100.0 & 0.0 & 0.0\\
3 & 0.202 & 0.072 & 0.090 & 0.322 & 0.059 & 0.054 & 0.877 & 0.851 & 63.9 & 100.0 & 6.1 & 0.7\\
4 & 0.109 & 0.027 & 0.098 & 0.121 & 0.048 & 0.055 & 0.881 & 0.864 & 89.5 & 100.0 & 43.7 & 31.9\\
5 & 0.052 & 0.007 & 0.005 & 0.031 & 0.018 & 0.036 & 0.982 & 0.979 & 100.0 & 100.0 & 99.9 & 99.9\\
6 & 0.073 & -0.001 & 0.010 & 0.018 & 0.017 & 0.035 & 0.985 & 0.982 & 100.0 & 100.0 & 100.0 & 100.0\\
7 & 0.077 & -0.006 & 0.014 & 0.011 & 0.019 & 0.036 & 0.981 & 0.977 & 100.0 & 100.0 & 100.0 & 100.0\\
8 & 0.085 & -0.011 & 0.027 & 0.009 & 0.022 & 0.038 & 0.975 & 0.968 & 100.0 & 100.0 & 100.0 & 99.5\\
9 & 0.082 & -0.014 & 0.030 & 0.004 & 0.022 & 0.038 & 0.970 & 0.966 & 100.0 & 100.0 & 100.0 & 100.0\\
\bottomrule
\end{tabular}}
\par\smallskip
{\footnotesize \emph{Note.} $K_{\mathrm{fit}}$ is the fitted number of factors
(true $K=5$). Cutoffs: $\mathrm{RMSEA}\le.06$, $\mathrm{SRMR}\le.10$,
$\mathrm{CFI}\ge.90$, $\mathrm{TLI}\ge.90$.}
\end{table*}

Over-factoring behaved differently. Beyond $K_{\mathrm{fit}}=5$, loading recovery
worsened slightly (RMSE $=.073$ at six factors) yet the absolute indices were
unchanged or marginally better ($\mathrm{CFI}_H=.985$). Because adding factors
cannot worsen covariance reproduction, the absolute diagnostics are sensitive to
under-factoring but effectively blind to over-factoring, a central reason
the relative criteria are needed. Two further patterns deserve emphasis. First,
$\mathrm{SRMR}_H$ was uninformative here: its largest mean value across all
under-factored models was only $.073$, well within the $.10$ cutoff, so it flagged
nothing, leaving $\mathrm{CFI}_H$ and $\mathrm{TLI}_H$ as the most discriminating
absolute indices. Second, the ability of
$\mathrm{CFI}_H$ and $\mathrm{TLI}_H$ to flag under-factoring weakened as factor
correlation increased: at $\phi=.25$ the model with one factor too few had
$\mathrm{CFI}_H=.845$, but at $\phi=.5$ it rose to $.918$ and passed the $.90$
cutoff, because highly correlated factors are nearly collinear and dropping one
barely harms covariance reproduction.

Figure~\ref{fig:fit-trend} shows this recovery directly across the design. Every metric
improves steeply up to $K_{\mathrm{fit}}=5$ and is flat or slightly better beyond it,
mirroring the absolute-fit trend, so under-factoring is by far the worst case in every
cell. The specification level $K_0$ matters mainly for the under-factored models:
supplying more required loadings sharply lowers their false-negative rate (the $K_0$
colors fan out most at $K_{\mathrm{fit}}<5$), whereas the true and over-factored models
are already near the floor. Recovery also worsens somewhat with stronger factor
correlation and improves with more items per factor, but these shifts are small beside
the dominant effect of getting $K$ right. Thus what the diagnostics flag as poor fit is
exactly what shows poor loading recovery, and both are driven by under-factoring rather
than by how much structure the analyst specifies.

\subsubsection{Relative selection of the number of factors}
\begin{figure*}[tp]
\centering
\includegraphics[width=\linewidth]{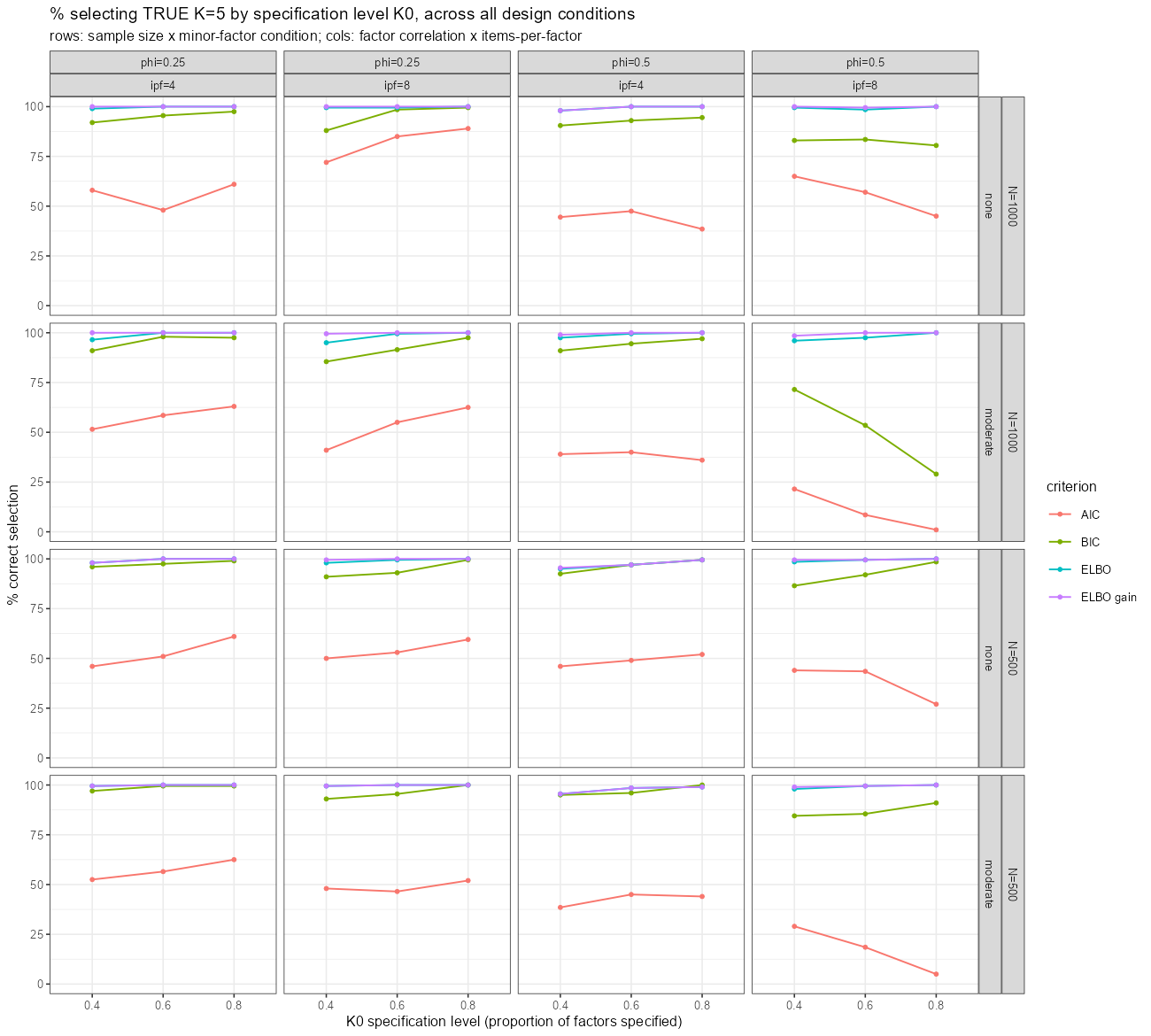}
\caption{Percentage of replications selecting the true $K$ as a function of the
specification level $K_0$, by criterion, across all design conditions.
AIC and BIC are shown as their hard variant.}
\label{fig:sel-k0}
\end{figure*}

Table~\ref{tab:rel-sel} and Figure~\ref{fig:sel-k0}
summarize how often each relative criterion recovered the true $K=5$ within the
candidate triplet $\{4,5,6\}$. No criterion under-selected ($K-1$ chosen less than
$1\%$ of the time); the difficulty lay entirely in distinguishing the true model
from the over-factored alternative. The ELBO recovered the true dimensionality
almost always ($99.0\%$), the two BIC variants were reliable
($\mathrm{BIC}_H=91.1\%$, $\mathrm{BIC}_S=91.9\%$), and the two AIC variants
over-factored roughly half the time ($\mathrm{AIC}_H=47.2\%$,
$\mathrm{AIC}_S=49.7\%$), reflecting AIC's weaker complexity penalty. Hard and soft
versions behaved almost identically. The scale-free gain rule
(Equation~\ref{eq:elbow}) applied to the ELBO recovered the true $K$ in $99.4\%$ of
replications (Table~\ref{tab:rel-sel}), and the same rule applied to the BIC and AIC
curves did almost as well ($98.7\%$ and $89.8\%$); the gain rule thus lifts the weakest
raw criterion, AIC, from $47\%$ to $90\%$. In this low-dimensional, low-nuisance regime
all three gain variants therefore work well and agree.

\begin{table*}[tp]
\centering
\caption{Relative fit: selection of the true factoring model.}
\label{tab:rel-sel}
\resizebox{\linewidth}{!}{%
\begin{tabular}[t]{lrrrrrrrr}
\toprule
$K_{\mathrm{fit}}$ & AIC$_H$ & AIC$_S$ & BIC$_H$ & BIC$_S$ & ELBO & AIC gain & BIC gain & ELBO gain\\
\midrule
Under & 0.0 & 0.0 & 0.1 & 0.1 & 0.3 & 0.1 & 0.1 & 0.3\\
True & 47.2 & 49.7 & 91.1 & 91.9 & 99.0 & 89.8 & 98.7 & 99.4\\
Over & 52.8 & 50.3 & 8.9 & 8.0 & 0.8 & 10.1 & 1.1 & 0.2\\
\bottomrule
\end{tabular}}
\par\smallskip
{\footnotesize \emph{Note.} Entries are the percentage of replications in which each
criterion (columns) selects the under-factored, true, or over-factored model, pooled over all conditions.}
\end{table*}

Figure~\ref{fig:sel-k0} relates selection accuracy to the specification level $K_0$ and
shows that this ordering was stable across the design, with informative exceptions. The
ELBO and the ELBO gain held near $100\%$ in every cell. Supplying more required loadings
(larger $K_0$) generally helped the information criteria (e.g., at $\mathrm{ipf}=8$,
$\phi=.25$, BIC rose from about $90\%$ to $99\%$), but in the most demanding cell (many
indicators and strong factor correlation, $\mathrm{ipf}=8$, $\phi=.5$) stronger
specification instead \emph{lowered} their accuracy (AIC from about $40\%$ to $20\%$;
BIC from about $82\%$ to $75\%$). A counterintuitive sample-size effect appeared in that
same cell: BIC selected the true model \emph{less} often at $N=1000$ than at $N=500$,
because with more information the extra factor begins to capture the genuine (if minor)
residual covariance, so the consistency property that normally favors the simpler model
instead rewards the larger one; the ELBO, whose penalty derives from the variational
Kullback--Leibler term rather than a fixed $\log N^{\ast}$ factor, was unaffected. The
minor-factor manipulation otherwise changed selection only modestly (by a few points for
the ELBO and BIC, and by $13$--$15$ points for AIC). Stronger partial specification is
therefore not a substitute for a robust selection criterion when the data contain
substantial nuisance structure.

\subsubsection{Summary}
Three conclusions follow. First, the absolute indices behave as intended for
\emph{confirming} an adequately specified major pattern: good recovery and good fit
coincide, and under-factoring is clearly flagged by $\mathrm{CFI}_H$ and
$\mathrm{TLI}_H$; but the indices cannot by themselves rule out over-factoring, and
$\mathrm{RMSEA}_H$ and $\mathrm{SRMR}_H$ add little in this setting. Second, for
\emph{choosing} among candidate dimensionalities the ELBO and its gain rule are the
most reliable and are uniquely robust to high factor correlation, minor structure,
larger samples, and the amount of supplied specification; BIC is a reasonable
secondary criterion, whereas AIC over-factors too often to be recommended. Third,
because retention rules disagree, the practical recommendation is to treat the
factor number as a range, recover the major pattern across that range, and rely on
the ELBO, supported by BIC and the absolute diagnostics, to identify the most
defensible solution, precisely the partially exploratory stance advocated in this
paper.

% (Study-1 figures are placed next to their text above.)

\subsection{Simulation Study Two}

Simulation Study Two scales the design to a high-dimensional regime that matches,
in scale, the empirical inventory analyzed below: $K=10$ major factors, $\mathrm{ipf}=8$
items per factor ($J=80$), $N=1000$, homogeneous $\phi=.25$, and major loadings
$.6$ with $\mathrm{cpf}=4$ alternating cross-loadings of $.2$. Nuisance is varied
through definition-based minor factors \parencite{auerswald2019determine}, each loading
on every item, at three increasing levels: \emph{1w+1m} (one weak, one moderate),
\emph{2w+2m} (two of each), and \emph{3w+3m} (three of each). Each data set is fit
over a window $K_{\mathrm{fit}}=K\pm3$ (i.e., $7$--$13$) at specification levels
$K_0\in\{.4,.6,.8\}K$, with $100$ replications per condition. All fits converged.

\subsubsection{Parameter recovery and absolute fit}
\begin{figure*}[tp]
\centering
\includegraphics[width=\linewidth]{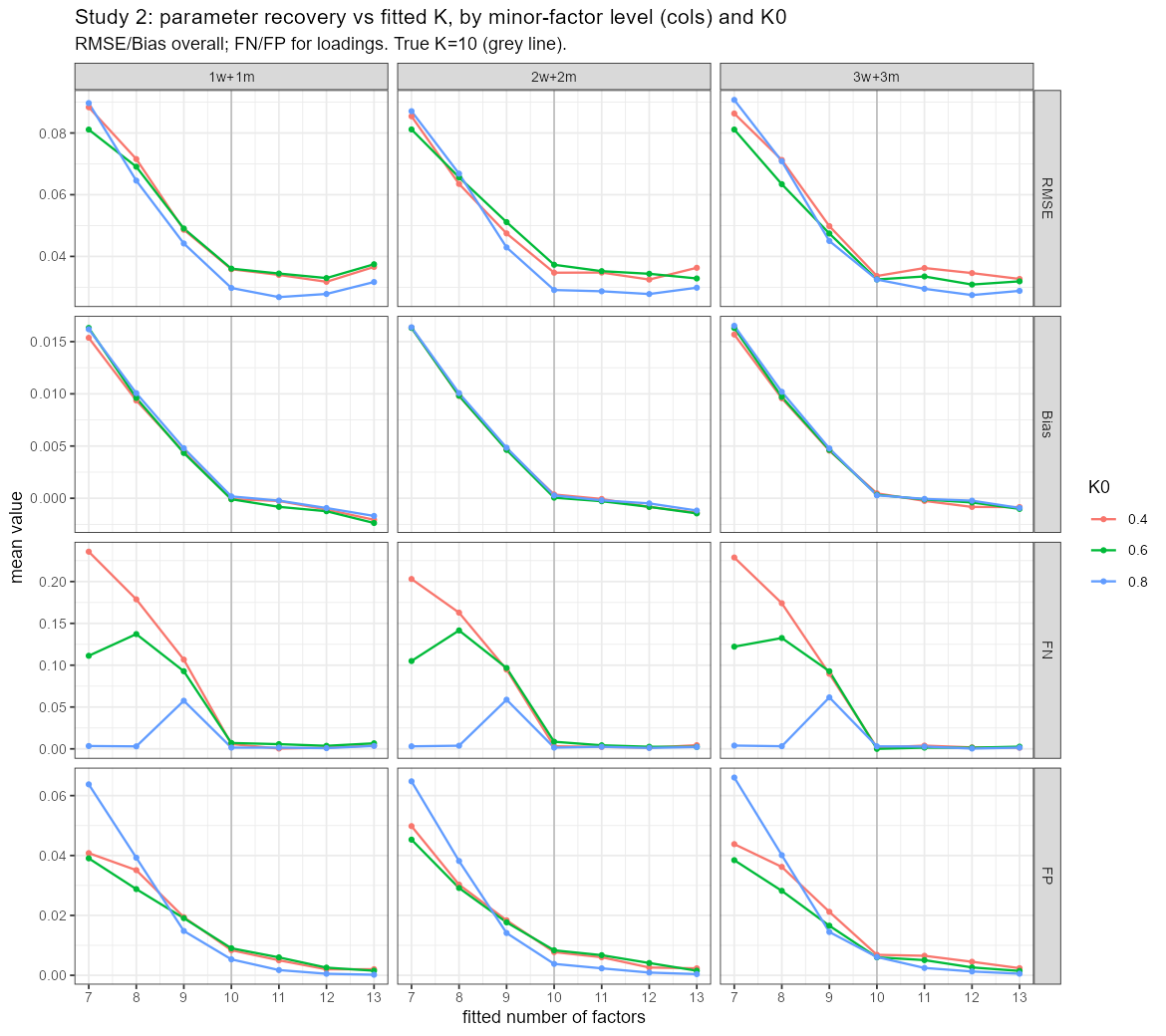}
\caption{Study 2: parameter recovery (overall RMSE and Bias; loading FN and FP)
versus the fitted number of factors, by minor-factor level (columns) and
specification level $K_0$ (color); sample size, items-per-factor, and $\phi$ are fixed
in Study 2, so these are all the varying design factors. True $K=10$ (grey line).
Recovery is good at the true $K$ and above and breaks only below it, in every cell.}
\label{fig:s2-rec}
\end{figure*}

Table~\ref{tab:s2-rec} reports parameter recovery and absolute fit as a function of
the fitted number of factors (pooled over nuisance level and $K_0$), and
Figure~\ref{fig:s2-rec} breaks the recovery out by those factors. \emph{Parameter
recovery is categorically different below versus at-or-above the truth.} For
$K_{\mathrm{fit}}<10$ it is poor and worsens as factors are dropped (overall RMSE
up to $.086$ and a false-negative rate of $.113$ at $K_{\mathrm{fit}}=7$); at
$K_{\mathrm{fit}}=10$ and at \emph{every} dimensionality above it, recovery is
excellent (RMSE $\approx.033$, FN $\le.004$, near-zero bias, FP $\le.007$). Notably,
over-factored models recover the major pattern \emph{as well as or slightly better
than} the true model (e.g., RMSE $.031$ and FP $.002$ at $K_{\mathrm{fit}}=12$),
because the extra factor absorbs the nuisance and leaves the major loadings
cleaner. Figure~\ref{fig:s2-rec} shows this under/at-or-above split holds in every
cell, across all three nuisance levels and all $K_0$.

Absolute fit shows the same asymmetry, but only $\mathrm{CFI}$ and $\mathrm{TLI}$
register it: their $.90$ cutoffs are met by $0\%$ of under-factored models and by
$100\%$ at $K_{\mathrm{fit}}\ge10$, so they cleanly separate under-factoring from
true-or-above factoring, though, like every absolute index, they cannot separate
over-factoring from true factoring (both pass). The conventional RMSEA $\le.06$ and
SRMR $\le.10$ cutoffs are \emph{insensitive to every model} (passing $100\%$ of
fits, even the worst under-factored ones) and are of no use at this scale.

\begin{table*}[tp]
\centering
\caption{Study 2: parameter recovery and absolute fit by the fitted number of
factors (true $K=10$), averaged over nuisance level and $K_0$.}
\label{tab:s2-rec}
\resizebox{\linewidth}{!}{%
\begin{tabular}[t]{crrrrrrrrrrrr}
\toprule
\multicolumn{1}{c}{ } & \multicolumn{4}{c}{Parameter recovery} & \multicolumn{4}{c}{Absolute fit (mean)} & \multicolumn{4}{c}{\% within cutoff} \\
\cmidrule(l{3pt}r{3pt}){2-5} \cmidrule(l{3pt}r{3pt}){6-9} \cmidrule(l{3pt}r{3pt}){10-13}
$K_{\mathrm{fit}}$ & RMSE & Bias & FN & FP & RMSEA & SRMR & CFI & TLI & RMSEA & SRMR & CFI & TLI\\
\midrule
7 & 0.086 & 0.016 & 0.113 & 0.050 & 0.044 & 0.061 & 0.740 & 0.728 & 100 & 100 & 0.0 & 0.0\\
8 & 0.067 & 0.010 & 0.104 & 0.034 & 0.038 & 0.055 & 0.807 & 0.798 & 100 & 100 & 0.0 & 0.0\\
9 & 0.047 & 0.005 & 0.083 & 0.017 & 0.031 & 0.047 & 0.876 & 0.870 & 100 & 100 & 7.2 & 2.9\\
10 & 0.033 & 0.000 & 0.004 & 0.007 & 0.020 & 0.034 & 0.949 & 0.946 & 100 & 100 & 100.0 & 100.0\\
11 & 0.033 & 0.000 & 0.003 & 0.005 & 0.018 & 0.033 & 0.956 & 0.953 & 100 & 100 & 100.0 & 100.0\\
12 & 0.031 & -0.001 & 0.002 & 0.002 & 0.018 & 0.033 & 0.957 & 0.954 & 100 & 100 & 100.0 & 100.0\\
13 & 0.033 & -0.001 & 0.003 & 0.001 & 0.018 & 0.034 & 0.955 & 0.952 & 100 & 100 & 100.0 & 100.0\\
\bottomrule
\end{tabular}}
\par\smallskip
{\footnotesize \emph{Note.} RMSE and Bias are overall (loadings and factor
correlations); FN and FP are false-negative/positive rates for the loadings. Cutoffs: RMSEA $\le.06$, SRMR $\le.10$,
CFI $\ge.90$, TLI $\ge.90$.}
\end{table*}

\subsubsection{Relative selection and the gain rule}
\begin{figure*}[tp]
\centering
\includegraphics[width=\linewidth]{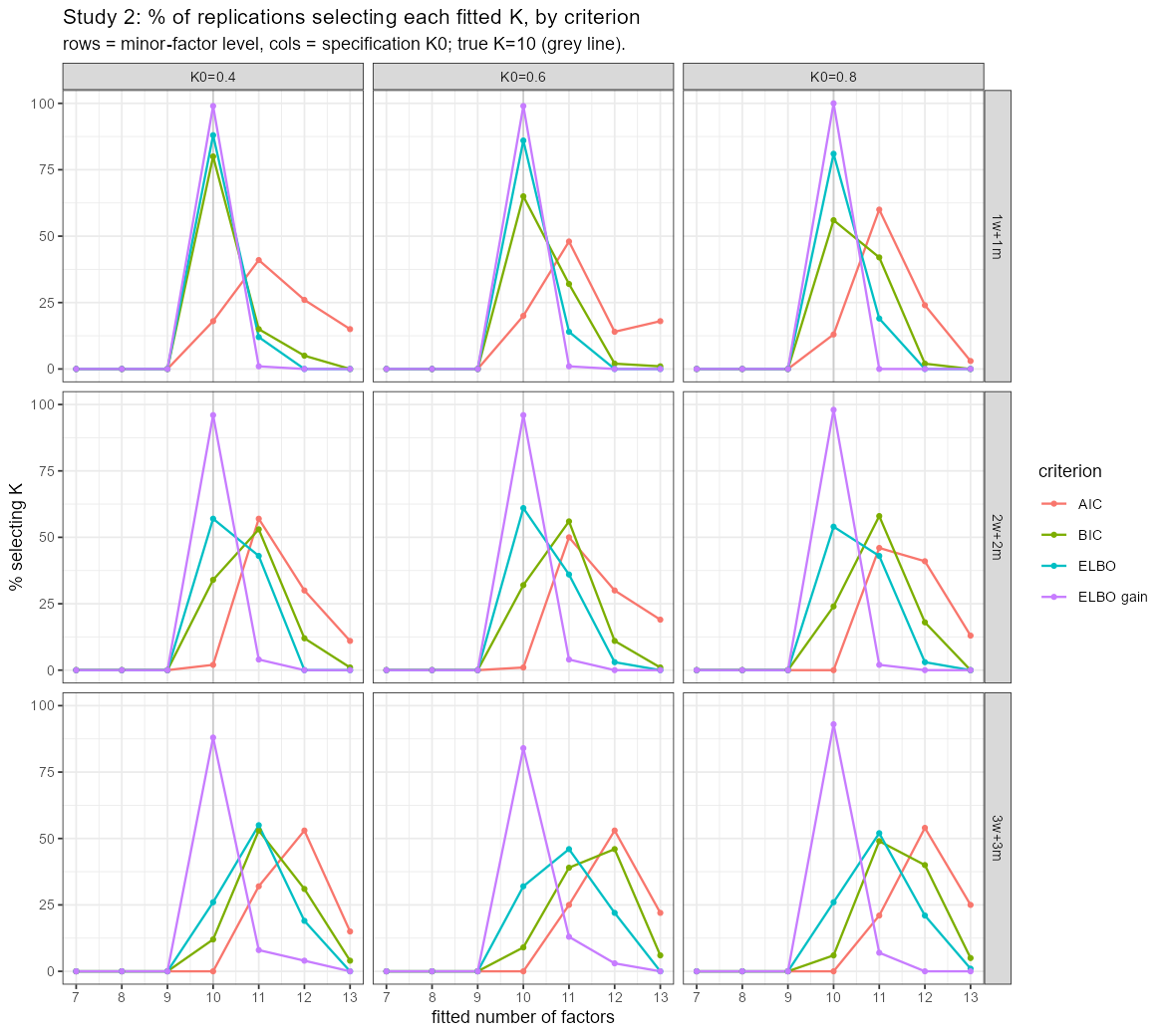}
\caption{Study 2: percentage of replications selecting each fitted $K$, by
criterion, with rows = minor-factor level and columns = specification level $K_0$.
AIC and BIC are shown as their hard variant (the soft variant is essentially
identical; see Table~\ref{tab:s2-sel}). True $K=10$ (grey line); the gain rule
concentrates at the truth while the raw criteria shift right as nuisance grows.}
\label{fig:s2-sel}
\end{figure*}

Table~\ref{tab:s2-sel} gives the distribution of the selected $K$ for each
criterion, and Figure~\ref{fig:s2-sel} breaks it out by nuisance and $K_0$. No
criterion ever selects $K<10$; the contest is between the true $K$ and
over-factoring, and the raw optima drift upward, increasingly so as nuisance grows.
Across the 1w+1m, 2w+2m, and 3w+3m levels the true $K$ is recovered by ELBO in
$85\%$, $57\%$, and $28\%$ of replications, by BIC in $67\%$, $30\%$, and $9\%$, and
by AIC in $17\%$, $1\%$, and $0\%$; accuracy also falls, counterintuitively, under
\emph{stronger} specification, for the same reason it falls with larger $N$ in
Study One: a richer backbone pins the major factors more firmly, freeing an added
factor to align with the genuine, if minor, nuisance covariance, which the raw
criteria then reward. The hard and soft variants of AIC and BIC are
essentially identical (Table~\ref{tab:s2-sel}), so Figure~\ref{fig:s2-sel} shows a
single representative of each. The ordering is thus
$\text{ELBO}>\text{BIC}>\text{AIC}$.

Applying the gain rule (with its sustained-drop guard, $s=2$) to each ELBO path
sharply improves on the raw optimum. It recovers the true $K$ in $95\%$ of
replications overall and in $99\%$, $97\%$, and $88\%$ across the three nuisance
levels, stays high for every $K_0$, and degrades only gently under the heaviest
nuisance, where every raw criterion has collapsed. The guard itself is nearly
free here: the first-crossing variant ($s=1$) is more accurate by less than one
point ($95.6\%$ versus $94.8\%$ overall), a small premium that buys the
protection against transient dips that the empirical analysis below requires. The ordering becomes
$\text{ELBO gain}>\text{ELBO}>\text{BIC}>\text{AIC}$. The $K\pm3$ window kept the
ELBO-gain selection strictly interior in every replication, so the boundary extension
described in the framework was never triggered; recovering the true number of factors
here is, in effect, what the gain rule is designed to do.

\begin{figure*}[tp]
\centering
\includegraphics[width=\linewidth]{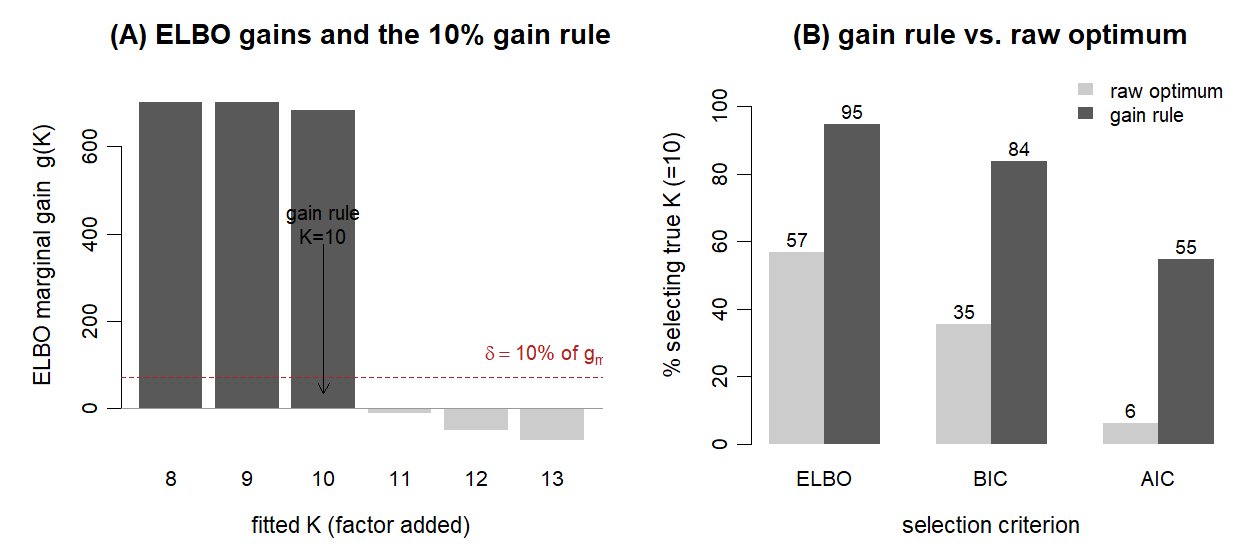}
\caption{Over-factoring is a property of the raw optimum, not of the criterion (means
and rates over all conditions). (A) ELBO marginal gains $g(K)$: the 8th--10th factors
each add $\approx700$ nats, then gains fall to $\le0$; the dashed line is the gain-rule
threshold $\delta=10\%$ of the largest gain, and the selected $K$ is the last factor
clearing it ($K=10$). (B) Percentage of replications selecting the true $K$ by the raw
optimum versus the gain rule, for each criterion; the same gain rule improves all three,
most so for the algorithm-native ELBO.}
\label{fig:s2-elbow}
\end{figure*}

Figure~\ref{fig:s2-elbow} shows that the over-factoring is a property of the raw
\emph{optimum}, not of the criterion. The mean ELBO marginal gains
(Figure~\ref{fig:s2-elbow}A) show the genuine 8th, 9th, and 10th factors each adding
roughly $700$ nats, after which the gain collapses to at or below zero. What over-factors
is the raw optimum, which keeps any factor with a positive marginal change and so chases
the tiny, noisy movements past the truth: the ELBO's negligible post-truth gains and,
symmetrically, the further small decreases of AIC and BIC (whose averaged minima sit at
$K=12$ and $K=11$). The gain rule instead keeps a factor only when its marginal change
exceeds $10\%$ of the dominant change, and this same rule applied to \emph{any} of
the three curves recovers the truth far more often than its raw optimum
(Figure~\ref{fig:s2-elbow}B): ELBO from $57\%$ to $95\%$, BIC from $35\%$ to $84\%$, and
AIC from $6\%$ to $55\%$. An AIC or BIC gain rule therefore helps substantially; the
ELBO gain is nonetheless the most accurate, and the ordering
$\text{ELBO}>\text{BIC}>\text{AIC}$ persists at every nuisance level, because the
estimated ELBO path is nondecreasing up to optimization noise with a single dominant
kink and is the algorithm-native quantity, evaluated without a plug-in covariance or
a rank-adjusted parameter count. We therefore adopt the ELBO gain, while noting that
the gain principle itself is general.

\begin{table*}[tp]
\centering
\caption{Study 2: percentage of replications selecting each fitted $K$, by
criterion (true $K=10$), averaged over nuisance level and $K_0$.}
\label{tab:s2-sel}
\resizebox{\linewidth}{!}{%
\begin{tabular}[t]{crrrrrrrr}
\toprule
$K_{\mathrm{fit}}$ & AIC$_H$ & AIC$_S$ & BIC$_H$ & BIC$_S$ & ELBO & AIC gain & BIC gain & ELBO gain\\
\midrule
7 & 0.0 & 0.0 & 0.0 & 0.0 & 0.0 & 0.0 & 0.0 & 0.0\\
8 & 0.0 & 0.0 & 0.0 & 0.0 & 0.0 & 0.0 & 0.0 & 0.0\\
9 & 0.0 & 0.0 & 0.0 & 0.0 & 0.0 & 0.0 & 0.0 & 0.0\\
10 & 6.0 & 5.9 & 35.3 & 34.8 & 56.8 & 54.7 & 83.7 & 94.8\\
11 & 42.2 & 42.2 & 44.1 & 45.2 & 35.6 & 34.1 & 12.4 & 4.4\\
12 & 36.1 & 36.9 & 18.6 & 18.1 & 7.6 & 10.0 & 3.7 & 0.8\\
13 & 15.7 & 15.0 & 2.0 & 1.9 & 0.1 & 1.2 & 0.2 & 0.0\\
\bottomrule
\end{tabular}}
\par\smallskip
{\footnotesize \emph{Note.} Each column is the percentage of replications whose
selected $K$ equals the row value (columns sum to $100$). Subscripts $H$/$S$ denote
the hard- and soft-selected variants, which are essentially identical. ELBO is the
raw optimum; the last three columns apply the scale-free gain rule ($\delta=10$,
sustained-drop guard $s=2$) to the AIC, BIC, and ELBO curves. No criterion selects $K<10$.}
\end{table*}

\subsubsection{Summary}
Two complementary uses follow. First, ELBO and BIC offer a useful guarantee even
when they over-factor: they \emph{never} under-select, and, comparing
Figures~\ref{fig:s2-rec} and \ref{fig:s2-sel}, because the chosen $K$ is always at
or above the truth, the selected model \emph{always} recovers the major loading
pattern, regardless of nuisance and $K_0$ (over the $900$ selections, mean overall
RMSE $.030$, mean FN $.003$). Any additional factor is therefore minor and may be
kept or discarded on substantive grounds. Moreover, a gap between the ELBO/BIC
optimum and the more parsimonious gain-rule solution ($\widehat{K}_{\mathrm{ELBO}}>
\widehat{K}_{\delta}$) is itself a useful signal that one or more minor factors are
present. Second, when a single parsimonious dimensionality is the goal, the
scale-free gain rule recovers the true $K$ almost always and is robust to
$N\cdot J$ scale, nuisance, and specification level, whereas the raw criteria
(especially AIC) are not.

\section{Empirical Example}

We illustrate the workflow on the Personality Inventory for DSM-5 (PID-5;
\textcite{krueger2012initial}), a maladaptive-trait questionnaire. We analyze its $100$-item Faceted Brief
Form, designed to measure $K=25$ narrow personality facets (four items each). The responses come from a French sample of
$N=2{,}532$ adults \parencite{roskam2015psychometric}. The
$25$-facet design serves only as an external benchmark; the loading structure is
treated as unknown at the specification stage.

\subsection{A window for the number of factors}
We first bracket the plausible number of factors with several retention rules.
Parallel analysis returns $16$ (both the mean-eigenvalue and the $95$th-percentile
variants), the empirical Kaiser criterion returns $15$, and the ordinary Kaiser
rule returns $21$. Two further rules gave degenerate values and were set aside as
implausible: the Hull method collapsed to a single factor, and the sequential model
test ran up to $46$. The credible estimates therefore span $15$ to $21$. To avoid
prejudging the dimensionality we adopt a deliberately wide window,
$K_{\mathrm{fit}}\in\{15,\dots,28\}$, that brackets these estimates, spans the $25$
designated facets, and leaves over-factoring headroom for the gain rule to locate an
interior elbow.

\subsection{A partially exploratory analysis}
We anchor a backbone $\mathbf{Q_0}$ on ten facets of the DSM-5 domains as ten factors, using each facet's two strongest-loading items as
markers and leaving every other loading unspecified (call this backbone~a). Sweeping
this backbone across the window and applying the ELBO gain rule with the
sustained-drop guard selects $\widehat{K}=22$; the raw ELBO optimum is $23$, and the
one-factor gap is the familiar over-factoring of the unpenalized bound, flagging a
single further minor factor. The selection is insensitive to the gain-rule
threshold: $\widehat{K}=22$ for every $\delta$ from $5$ to $20$ under backbone~a
(and from $5$ to $30$ under the disjoint backbone~b introduced below), with only
extreme values moving it, as $\delta=2$ admits a noise-level gain and $\delta\geq25$
prunes the borderline factors (Appendix~\ref{app:gainrule}, Table~\ref{tab:app-delta-emp}).
The full sweep is in Appendix Table~\ref{tab:app-sweep}.
The selected solution fits well, and better than the confirmatory model
(Table~\ref{tab:emp-fit}): $\operatorname{RMSEA}_H=.029$, $\operatorname{SRMR}_H=.051$,
$\operatorname{CFI}_H=.910$, and $\operatorname{TLI}_H=.901$, each improving on the
25-facet CFA at a comparable parameter count, with the confirmatory model better only on
$\operatorname{SRMR}$ (it retains all $100$ simple-structure loadings). This comparison
is conservative for the partially exploratory rows: their statistics are plug-in
quantities evaluated at the shrunken variational posterior means, which bound the
minimized discrepancy of the selected pattern from above (see the framework section),
whereas the CFA row is evaluated at \texttt{lavaan}'s exact ML optimum. Refitting the
hard-selected pattern by ML could therefore only widen the reported margins; the
partially exploratory models win despite conceding this handicap. The recovered factors are also
far more distinct than the designed facets: their largest correlation is $.55$ with
no pair above $.70$, whereas the 25 confirmatory facets correlate up to $.94$ with five
pairs above $.70$, the long-noted discriminant-validity problem of the PID-5.
Matching each recovered factor to its most congruent designed facet, seven of the
twenty-five facets are recovered as distinct, one-to-one factors at Tucker's
congruence $\ge.90$; the remaining factors are interpretable blends of collinear facets.

\begin{table*}[tp]
\centering
\caption{Empirical example: fit of two partially exploratory models
against the confirmatory $25$-facet model.}
\label{tab:emp-fit}
\begin{tabular}[t]{lrrrrrrrr}
\toprule
Model & $t$ & RMSEA & SRMR & CFI & TLI & AIC & BIC & $\max|\phi|$\\
\midrule
PEFA (a), $K{=}22$ & 529 & .029 & .051 & .910 & .901 & 619{,}804 & 622{,}891 & .55\\
PEFA (b), $K{=}22$ & 533 & .029 & .051 & .909 & .901 & 619{,}826 & 622{,}937 & .59\\
CFA, $K{=}25$      & 500 & .034 & .049 & .880 & .869 & 623{,}173 & 626{,}091 & .94\\
\bottomrule
\end{tabular}
\par\smallskip
{\footnotesize \emph{Note.} The two partially exploratory PID-5 models use
	\emph{disjoint} ten-factor backbones (a and b, hard selection at $\tau=.50$, $K=22$); $N=2{,}532$; $t$ = free parameters,
	$\max|\phi|$ = largest factor correlation. Partially exploratory rows are the
	hard-selection SEM-like quantities; the CFA row is \texttt{lavaan}'s native
	maximum-likelihood fit to the standardized items.}
\end{table*}

\subsection{Robustness to the specification}
The choice of which ten facets to anchor has no strong a priori justification, so the
decisive question for a partially exploratory analysis is whether the recovered
structure depends on that choice. We test this in the sharpest possible way, with a
second backbone~b that anchors ten \emph{entirely different} facets, zero overlap with
a, under the identical design (two markers each). Were the method at the mercy of the
specification, a and b would diverge. They do not, where the theory says they should
not. Backbone~b selects the same
$\widehat{K}=22$ (its ELBO gain dips transiently at $K=21$ and rebounds, so the
sustained-drop guard, not a single sub-threshold step, sets the elbow; Appendix
Table~\ref{tab:app-sweep}), fits essentially identically (Table~\ref{tab:emp-fit}),
and its hard-selected loading matrix is congruent with that of~a at mean $.90$,
with twelve of the twenty-two factors matching at $\ge.90$. The agreement is
concentrated exactly where it should be: the factors that coincide are the strong,
well-separated dimensions, whereas the factors that differ lie in the mutually
collinear sub-block whose internal resolution, as the next paragraphs show, is
precisely what the specification controls. The major latent pattern is
thus recovered stably no matter which reasonable set of facets is anchored, while
agreement within the entangled sub-block is not claimed and should not be expected.

What the specification does control is the resolution of \emph{individual, mutually
collinear} facets (Table~\ref{tab:emp-flip}). Attention Seeking is the clearest case:
a does not anchor it, and it merges with Separation Insecurity into one factor
(congruence $.65$); b anchors both, and each separates cleanly (congruence $1.00$ and
$.98$). The effect is local and symmetric, facets that a anchors and b leaves free
move the other way (e.g., Emotional Lability), while well-separated facets are
recovered identically by both whether anchored or not (Hostility and Intimacy
Avoidance, congruence $\ge.95$ in both). Averaged over all facets the anchoring effect
is small: a facet is recovered only about $.05$ better when the model anchors it,
because most factors are strong, well-separated data dimensions that neither backbone
can disturb. The movement is confined to the collinear sub-block. Anchoring is,
moreover, necessary but not sufficient: Suspiciousness ($.58$) and Perceptual
Dysregulation ($.55$) are anchored by backbone~b yet recovered poorly under both
backbones (Appendix Table~\ref{tab:app-facetrec}), because their items load
diffusely across several recovered dimensions, so no two-marker anchor can pull
them into a distinct factor; resolving such facets would require either stronger
anchors or a substantive decision that they are not distinct dimensions in these
data.

\begin{table}[t]
\centering
\caption{Empirical example: the specification controls only the resolution of
collinear facets. For selected facets, the maximum Tucker congruence between the
designed facet and any recovered factor, under backbone~a and the disjoint backbone~b,
with the backbone (if any) that anchors the facet. Full table in
Appendix Table~\ref{tab:app-facetrec}.}
\label{tab:emp-flip}
\begin{tabular}[t]{llrr}
\toprule
Facet & anchored by & in a & in b\\
\midrule
Attention Seeking      & b       & .65 & 1.00\\
Separation Insecurity  & b       & .76 & .98\\
Risk Taking            & b       & .86 & 1.00\\
Callousness            & neither & .72 & .95\\
Emotional Lability     & a       & .94 & .80\\
Restricted Affectivity & neither & .76 & .52\\
Hostility              & b       & .95 & .95\\
Intimacy Avoidance     & b       & 1.00 & 1.00\\
\midrule
\multicolumn{2}{l}{facets recovered at $\ge.90$ (of $25$)} & 7 & 12\\
\bottomrule
\end{tabular}
\end{table}

This delineates the identifiable target of the method. The \emph{major structure},
the strong, well-separated factors, is recovered robustly and is not a function of the
backbone. The internal composition of a cluster of mutually collinear factors, by
contrast, is identified only when the backbone anchors within that cluster; a cluster
left entirely unspecified is combined in a specification-dependent way. The
implication is not fragility but control: the researcher resolves the entangled facets
through the backbone, and anchoring more, or anchoring inside a collinear cluster of
substantive interest, resolves more of them. A richer specification bears this out. A
fifteen-facet backbone (three per domain) resolves thirteen of the
twenty-five facets as distinct factors, against the ten-facet backbone's seven, while selecting the same $K$ region and returning the same major structure. In
a genuinely partially exploratory study one would therefore choose the backbone to
anchor the clusters of substantive interest, from theory where it exists or, absent
it, from a content-based reading of the items.

\section{Discussion}

This paper equips the Bayesian variable-selection machinery of the regularized variational approximation for partially confirmatory factor analysis with post-selection fit assessment and a factor-number selection rule, so that the same machinery can be used in partially exploratory settings, where the loading structure and the number of factors $K$ are only weakly specified. The central device is to convert the converged PCFA-VA solution into a covariance model in one of two ways. Hard selection thresholds the variational inclusion probabilities $\widehat{\pi}_{jk}$ at $\tau$ to produce a fixed sparse loading pattern, which behaves like a conventional sparse CFA/SEM model and supports the absolute diagnostics $\operatorname{RMSEA}_{H}$, $\operatorname{SRMR}_{H}$, $\operatorname{CFI}_{H}$, and $\operatorname{TLI}_{H}$ together with the relative criteria $\operatorname{AIC}_{H}$ and $\operatorname{BIC}_{H}$. Soft selection retains the inclusion probabilities as effective loading weights; its parameter count is the variational posterior expectation of the number of active loadings, an effective number of parameters in the tradition of $p_D$, $p_{\mathrm{WAIC}}$, and generalized degrees of freedom, and the hard--soft gap serves as a built-in diagnostic for how much selection uncertainty the thresholding discards. For choosing $K$ across a window of candidates, the scale-free gain rule with its sustained-drop guard thresholds the marginal gain of a criterion against its largest gain, provably recovers a bracketed truth whenever the gains separate (Proposition~1 in Appendix~\ref{app:gainrule}), and so avoids the over-factoring to which the raw optima are prone.

The simulations map out when these tools work. The absolute indices behave as intended for \emph{confirming} an adequately specified major pattern: good recovery and good fit coincide, and $\operatorname{CFI}_{H}$ and $\operatorname{TLI}_{H}$ clearly flag under-factoring; but no absolute index can rule out over-factoring, which is why a relative criterion is needed. Among the relative criteria the raw optima over-factor, mildly for the ELBO and BIC and severely for AIC, and increasingly so as nuisance structure grows; the gain rule corrects this, lifting even AIC to usable accuracy, and the ELBO gain, computed on the algorithm-native bound, is the most robust variant under heavy nuisance and high factor correlation.

The empirical example carries these findings to real data and confronts the central objection to any partially exploratory method, that the answer might merely reflect the researcher's specification. Fitting the 100-item PID-5 with a ten-facet backbone selected $22$ factors and fit better than the confirmatory $25$-facet model on every incremental and parsimony-adjusted index, despite a plug-in evaluation that handicaps the partially exploratory solution, and with far fewer strongly correlated factor pairs. More tellingly, a second backbone that anchored ten \emph{entirely different} facets recovered the same dimensionality and the same major structure (mean loading congruence $.90$, with the disagreement confined to the mutually collinear sub-block), so the major latent pattern is not an artifact of the specification. What the specification does control is local: the resolution of a cluster of mutually collinear factors is identified only when the backbone anchors within that cluster, and a cluster left entirely unspecified is combined in a specification-dependent way. This delineates the identifiable target of the method. The strong, well-separated major structure is recovered robustly, whereas the internal composition of an entangled sub-block is resolved at the analyst's discretion, by anchoring within it.

For practice we recommend the hard-selected indices as the primary basis for reporting, the soft-selected quantities as a sensitivity analysis, and the ELBO gain as the primary criterion for comparing candidate values of $K$ or competing $\mathbf{Q}$ matrices, with BIC and the absolute diagnostics in support, all computed under identical data preprocessing, priors, and hyperparameters. Because retention rules disagree and the recovered structure is most trustworthy at the major level, the number of factors is best treated as a range rather than a point, with the goal of assessing whether the major pattern is stable and interpretable across it. When individual entangled facets are of substantive interest, the backbone should be chosen to anchor within their clusters, drawing on theory where it exists or, absent it, on a content-based reading of the items. Because all chi-square-type quantities are evaluated at regularized variational estimates after loading-pattern selection, they should be read as pseudo statistics and conventional cutoffs as descriptive heuristics rather than strict decision rules.

Several limitations bound these conclusions and point to extensions. The fit indices are plug-in quantities evaluated at a data-selected pattern rather than exact likelihood-ratio statistics: the plug-in direction is conservative (the shrunken variational estimates bound the pattern's minimized discrepancy from above), but the selection direction is optimistic, because the hard-selected degrees of freedom are conditional on a pattern the data chose, and formal post-selection inference for this two-sided bias remains open \parencite{berk2013valid,efron2004estimation}; the hard--soft parameter-count gap proposed here is a diagnostic for it, not a correction. Relatedly, the sampling behavior of the pseudo chi-square quantities, and any recalibration of cutoffs for the one-stage regularized setting, deserves further study. For the gain rule, Proposition~1 (Appendix~\ref{app:gainrule}) gives the exact bracketing-and-separation condition and Appendix~\ref{app:gainrule} shows the selections are stable over $\delta\in[5,20]$ in every setting examined, but automating the window choice would make the procedure more turnkey. The ELBO comparisons assume a comparable variational gap across candidates, which is untestable directly, although the simulations show the resulting selections are accurate where the truth is known. The identifiability boundary characterized here, that a collinear cluster left unspecified is combined arbitrarily, is a property of partially exploratory estimation in general and merits a formal treatment. Finally, the framework was developed for continuous, normal-theory covariance structures, and the empirical items are four-category approximations at the boundary of what that treatment supports \parencite{rhemtulla2012when}; extending the hard- and soft-selected indices to ordinal and categorical data, and to multiple-group and longitudinal designs, would broaden its applicability.

\section{Data and Code Availability}
The complete R implementation of the proposed fit indices and gain rule, together
with the simulation results behind every reported table (complete summaries for
Study One and per-replication results for Study Two, sufficient to verify the
threshold-sensitivity appendix without re-running any estimation) and the
empirical result files, is available for masked
review through an anonymized OSF view-only link:
\url{https://osf.io/q3bg2/overview?view_only=02dc13412c5e4a78860aa096cc74828c}.
The project will be made public and citable upon acceptance. Because the Personality Inventory for DSM-5 item
wording is copyrighted by the American Psychiatric Association and the response data of
\textcite{roskam2015psychometric} are subject to that source's terms, the original
empirical data are not redistributed; in their place the package provides a simulated
dataset with the same $100$-item, $25$-facet covariance structure that runs the full
empirical pipeline, so that all analyses can be verified without the original data.

\printbibliography

\appendix

\section{The gain rule: correctness condition and threshold sensitivity}
\label{app:gainrule}

\subsection{A correctness condition}
The gain rule of Equation~\ref{eq:elbow}, with the sustained-drop guard, admits an
exact finite-sample characterization of when it returns the true dimensionality.
Fix a window $\{K_{\min},\dots,K_{\max}\}$, let $g(K)=c(K)-c(K-1)$ for
$K=K_{\min}+1,\dots,K_{\max}$ denote the observed gains, let
$g_{\max}=\max_{K}g(K)>0$, and write $t_{\delta}=\tfrac{\delta}{100}\,g_{\max}$ for
the threshold. The guarded rule with parameter $s\geq1$ returns
\begin{equation}
\widehat{K}_{\delta,s}
=\min\bigl\{K:\ g(K+1)<t_{\delta},\ \dots,\ g(K+s)<t_{\delta}\bigr\},
\end{equation}
with windows truncated at $K_{\max}$ and $\widehat{K}_{\delta,s}=K_{\max}$ when no
such $K$ exists.

\begin{quote}
\textbf{Proposition 1.}
Suppose the window brackets the truth with $s$ steps of headroom,
$K_{\min}\leq K^{\ast}\leq K_{\max}-s$, and the gains \emph{separate} at
$K^{\ast}$:
(i) $g(K)\geq t_{\delta}$ for all $K_{\min}<K\leq K^{\ast}$, and
(ii) $g(K)<t_{\delta}$ for $K=K^{\ast}+1,\dots,K^{\ast}+s$.
Then $\widehat{K}_{\delta,s}=K^{\ast}$.

\emph{Proof.} For any candidate $K<K^{\ast}$, the run
$g(K+1),\dots,g(K+s)$ contains $g(K+1)$ with $K+1\leq K^{\ast}$, which by (i) is
at least $t_{\delta}$, so the run is not entirely sub-threshold and $K$ does not
qualify. For $K=K^{\ast}$, the run $g(K^{\ast}+1),\dots,g(K^{\ast}+s)$ lies within
the window by the headroom assumption and is entirely sub-threshold by (ii), so
$K^{\ast}$ is the smallest qualifying candidate. $\square$
\end{quote}

The conditions concern the \emph{estimated} gains, so the proposition converts the
selection problem into an empirical question: how often, and over how wide a range
of $\delta$, do fitted gain paths separate at the truth? Genuine factors reduce the
population discrepancy by fixed amounts while nuisance factors contribute gains
that vanish relative to $g_{\max}$, so separation holds whenever sampling and
optimization noise is small against the dominant gains; the tables below measure
this margin directly. Condition (ii) can fail for a \emph{single} step yet hold on
the next when a local variational optimum produces a transient dip, which is
exactly the failure the guard absorbs (backbone~b in
Appendix~\ref{app:sweeps} provides an observed instance).

\subsection{Sensitivity to the threshold \texorpdfstring{$\delta$}{delta}}
Table~\ref{tab:app-delta-sim} recomputes the gain-rule selections of both
simulation studies for $\delta$ between $2$ and $40$ (all with $s=2$), using the
saved criterion paths of every replication. Accuracy is high and flat over a wide
range: the ELBO gain stays above $94\%$ for every $\delta\in[10,40]$ in both
studies, and even the weakest criterion, the AIC gain, reaches the high
nineties once $\delta\geq20$. Small thresholds ($\delta\leq5$) admit noise-level
gains under heavy nuisance, which is the one regime where accuracy drops
materially. In these simulations larger $\delta$ can only help because every
sub-truth gain is enormous, so under-selection is impossible; the empirical sweeps
(Table~\ref{tab:app-delta-emp}) show the countervailing risk, namely that a large
threshold prunes genuine borderline factors ($\delta\geq25$ moves backbone~a from
$22$ to $20$ factors). The default $\delta=10$ therefore sits at the
under-selection-averse end of the stable region, and the selections reported in
the paper are unchanged for every $\delta\in[5,20]$ in both studies and both
backbones.

\begin{table*}[tp]
\centering
\footnotesize
\caption{Percentage of replications selecting the true $K$ by the gain rule
($s=2$) as a function of the threshold $\delta$, from the saved criterion paths
(Study 1: $9{,}600$ triplets, true $K=5$; Study 2: $900$ paths, true $K=10$).}
\label{tab:app-delta-sim}
\begin{tabular}[t]{lrrrrrrrr}
\toprule
& \multicolumn{8}{c}{$\delta$ (\% of largest gain)}\\
\cmidrule(l{3pt}r{3pt}){2-9}
Criterion & 2 & 5 & 10 & 15 & 20 & 25 & 30 & 40\\
\midrule
\multicolumn{9}{l}{\emph{Study 1 (true $K=5$)}}\\
ELBO gain & 99.2 & 99.3 & 99.4 & 99.5 & 99.5 & 99.6 & 99.6 & 99.6\\
BIC gain  & 94.4 & 96.9 & 98.7 & 99.3 & 99.5 & 99.6 & 99.6 & 99.7\\
AIC gain  & 63.2 & 78.4 & 89.8 & 95.3 & 98.0 & 99.1 & 99.5 & 99.8\\
\multicolumn{9}{l}{\emph{Study 2 (true $K=10$)}}\\
ELBO gain & 69.0 & 83.3 & 94.8 & 97.9 & 99.6 & 99.9 & 100.0 & 100.0\\
BIC gain  & 41.3 & 61.2 & 83.7 & 95.6 & 98.7 & 99.9 & 100.0 & 100.0\\
AIC gain  & 10.1 & 24.7 & 54.7 & 76.2 & 91.4 & 97.0 & 99.6 & 100.0\\
\bottomrule
\end{tabular}
\par\smallskip
{\footnotesize \emph{Note.} By-nuisance ELBO-gain accuracy in Study 2 at
$\delta=10$: $99.3$, $96.7$, and $88.3$ across the 1w+1m, 2w+2m, and 3w+3m
levels.}
\end{table*}

\begin{table*}[tp]
\centering
\footnotesize
\caption{Empirical example: the number of factors selected by the ELBO gain rule
($s=2$) as a function of $\delta$, for the two disjoint ten-facet backbones.}
\label{tab:app-delta-emp}
\begin{tabular}[t]{lrrrrrrrr}
\toprule
& \multicolumn{8}{c}{$\delta$ (\% of largest gain)}\\
\cmidrule(l{3pt}r{3pt}){2-9}
Backbone & 2 & 5 & 10 & 15 & 20 & 25 & 30 & 40\\
\midrule
a & 23 & 22 & 22 & 22 & 22 & 20 & 20 & 20\\
b & 25 & 22 & 22 & 22 & 22 & 22 & 22 & 20\\
\bottomrule
\end{tabular}
\end{table*}

\section{Number-of-factors sweeps for the two backbones}
\label{app:sweeps}
Table~\ref{tab:app-sweep} reports the full per-$K$ sweep over the window
$K_{\mathrm{fit}}=15$--$28$ for backbones a and b: the ELBO, its marginal gain, that
gain as a percentage of the largest gain in the window (the quantity the gain rule
thresholds at $10\%$), and the hard-selection $\operatorname{BIC}_H$,
$\operatorname{RMSEA}_H$, $\operatorname{CFI}_H$, and $\operatorname{TLI}_H$. Both
backbones select $\widehat{K}=22$ under the sustained-drop rule. Backbone b
illustrates the guard: its gain falls to $5\%$ at $K=21$ and rebounds to $32\%$ at
$K=22$, so a naive first-crossing rule would stop prematurely at $K=20$, whereas
requiring the drop to be sustained correctly returns $K=22$.

\begin{table*}[tp]
\centering
\footnotesize
\setlength{\tabcolsep}{4pt}
\caption{Per-$K$ sweep statistics for the two disjoint ten-facet backbones (hard
selection, $\tau=.50$). \%\,gain is the ELBO gain as a percentage of the largest gain
in the window; the gain-rule threshold is $10\%$.}
\label{tab:app-sweep}
\begin{tabular}[t]{rrrrrrrr}
\toprule
$K$ & ELBO & gain & \%\,gain & $\operatorname{BIC}_H$ & RMSEA & CFI & TLI\\
\midrule
\multicolumn{8}{l}{\emph{Backbone a: ten facets ($\widehat{K}=22$)}}\\
15 & $-313{,}321$ &      &     & 627{,}723 & .036 & .860 & .850\\
16 & $-312{,}812$ &  509 & 100 & 628{,}116 & .037 & .856 & .845\\
17 & $-312{,}437$ &  375 &  74 & 625{,}807 & .034 & .878 & .869\\
18 & $-312{,}111$ &  326 &  64 & 625{,}512 & .033 & .882 & .872\\
19 & $-311{,}726$ &  385 &  76 & 624{,}695 & .032 & .889 & .880\\
20 & $-311{,}433$ &  293 &  58 & 624{,}350 & .032 & .893 & .884\\
21 & $-311{,}330$ &  104 &  20 & 623{,}951 & .031 & .898 & .889\\
22 & $-311{,}226$ &  103 &  20 & 622{,}876 & .029 & .910 & .901\\
23 & $-311{,}213$ &   13 &   3 & 622{,}506 & .029 & .914 & .906\\
24 & $-311{,}417$ & $-204$ & $-40$ & 622{,}786 & .029 & .915 & .905\\
25 & $-311{,}490$ & $-72$ & $-14$ & 623{,}588 & .030 & .909 & .898\\
26 & $-311{,}487$ &    3 &   1 & 623{,}102 & .029 & .913 & .903\\
27 & $-311{,}392$ &   95 &  19 & 623{,}343 & .029 & .911 & .901\\
28 & $-311{,}594$ & $-201$ & $-40$ & 623{,}386 & .029 & .913 & .902\\
\multicolumn{8}{l}{\emph{Backbone b: ten disjoint facets ($\widehat{K}=22$)}}\\
15 & $-313{,}417$ &      &     & 628{,}110 & .037 & .857 & .846\\
16 & $-312{,}967$ &  450 &  92 & 627{,}399 & .036 & .865 & .854\\
17 & $-312{,}477$ &  491 & 100 & 626{,}241 & .034 & .874 & .864\\
18 & $-312{,}137$ &  340 &  69 & 625{,}652 & .034 & .880 & .870\\
19 & $-311{,}851$ &  286 &  58 & 625{,}683 & .033 & .881 & .871\\
20 & $-311{,}401$ &  450 &  92 & 623{,}587 & .031 & .900 & .892\\
21 & $-311{,}375$ &   26 &   5 & 623{,}680 & .031 & .901 & .892\\
22 & $-311{,}220$ &  155 &  32 & 622{,}906 & .029 & .910 & .901\\
23 & $-311{,}443$ & $-224$ & $-46$ & 623{,}633 & .030 & .906 & .896\\
24 & $-311{,}422$ &   21 &   4 & 622{,}792 & .029 & .915 & .905\\
25 & $-311{,}322$ &  100 &  20 & 622{,}591 & .028 & .916 & .907\\
26 & $-311{,}454$ & $-133$ & $-27$ & 622{,}776 & .029 & .916 & .906\\
27 & $-311{,}452$ &    2 &   0 & 623{,}070 & .029 & .914 & .904\\
28 & $-311{,}458$ & $-6$ & $-1$ & 623{,}646 & .030 & .909 & .898\\
\bottomrule
\end{tabular}
\end{table*}

\section{Facet recovery under the two disjoint backbones}
Table~\ref{tab:app-facetrec} gives, for every one of the $25$ designed facets, the
maximum Tucker congruence between that facet and any recovered factor, under backbone
a and under the disjoint backbone b, together with the backbone (if any) that anchors
the facet. Facets recovered congruently by both are the backbone-robust major
structure; the facets that move are the entangled ones, and they move toward whichever
backbone anchors them.

\begin{table}[t]
\centering
\footnotesize
\caption{Maximum facet-to-factor Tucker congruence under the two disjoint backbones.
$\Delta = $ (in b) $-$ (in a).}
\label{tab:app-facetrec}
\begin{tabular}[t]{llrrr}
\toprule
Facet & anchored by & in a & in b & $\Delta$\\
\midrule
Attention Seeking        & b       & .65 & 1.00 & $+.35$\\
Callousness              & neither & .72 & .95 & $+.23$\\
Separation Insecurity    & b       & .76 & .98 & $+.22$\\
Risk Taking              & b       & .86 & 1.00 & $+.14$\\
Impulsivity              & a       & .86 & 1.00 & $+.14$\\
Withdrawal               & a       & .79 & .92 & $+.13$\\
Depressivity             & neither & .64 & .65 & $+.01$\\
Distractibility          & b       & .75 & .75 & $\phantom{+}.00$\\
Rigid Perfectionism      & neither & .73 & .74 & $+.01$\\
Eccentricity             & a       & .96 & .96 & $\phantom{+}.00$\\
Grandiosity              & b       & .96 & .96 & $\phantom{+}.00$\\
Hostility                & b       & .95 & .95 & $\phantom{+}.00$\\
Intimacy Avoidance       & b       & 1.00 & 1.00 & $\phantom{+}.00$\\
Manipulativeness         & a       & .89 & .89 & $\phantom{+}.00$\\
Perseveration            & neither & .90 & .90 & $\phantom{+}.00$\\
Suspiciousness           & b       & .58 & .58 & $\phantom{+}.00$\\
Unusual Beliefs \& Exp.  & a       & .81 & .81 & $\phantom{+}.00$\\
Perceptual Dysregulation & b       & .55 & .55 & $-.01$\\
Anhedonia                & a       & .73 & .68 & $-.05$\\
Deceitfulness            & a       & .75 & .66 & $-.09$\\
Irresponsibility         & a       & .46 & .37 & $-.09$\\
Submissiveness           & b       & 1.00 & .91 & $-.09$\\
Anxiousness              & a       & .86 & .75 & $-.11$\\
Emotional Lability       & a       & .94 & .80 & $-.14$\\
Restricted Affectivity   & neither & .76 & .52 & $-.24$\\
\bottomrule
\end{tabular}
\end{table}

\end{document}